\documentclass[usenatbib]{mn2e}
\usepackage{epsfig}
\usepackage{times}
\usepackage{lscape}
\usepackage{multirow}
\usepackage{multicol}
\usepackage{subfigure}
\usepackage{color}
\usepackage{amsfonts}
\usepackage{amssymb}

\begin{document}

\title[X-ray reflection in GX 339$-$4] 
{Revealing accretion onto black holes: X-ray reflection throughout three outbursts of GX 339$-$4}
\author[Plant et al.] 
{D.~S. Plant$^1$\thanks{E-mail: d.plant@soton.ac.uk}, R.~P. Fender$^{2,1}$, G. Ponti$^{3}$, T. Mu\~{n}oz-Darias$^{2,1}$ and M. Coriat$^4$\\
$^1$School of Physics and Astronomy, University of Southampton, Highfield, Southampton, SO17 1BJ, UK\\
$^2$University of Oxford, Department of Physics, Astrophysics, Keble Road, Oxford, OX1 3RH, United Kingdom\\
$^3$Max Planck Institute fur Extraterrestriche Physik, 85748, Garching, Germany\\
$^4$Department of Astronomy, University of Cape Town, Private Bag X3, Rondebosch 7701, South Africa\\
\\
}
\maketitle

%%%%%%%%%%%%%%%%%%%%%%%%%%%%%%%%%%%%%%%%%%%%%%%%%%%%%%%%%%%%%%%%%%%%%%%%%%%%%%%%%%%%%%%%%

\begin{abstract}
Understanding the dynamics behind black hole state transitions and the changes they reflect in outbursts has become long-standing problem. The X-ray reflection spectrum describes the interaction between the hard X-ray source (the power-law continuum) and the cool accretion disc it illuminates, and thus permits an indirect view of how the two evolve. We present a systematic analysis of the reflection spectrum throughout three outbursts (500+ observations) of the black hole binary GX 339$-$4, representing the largest study applying a self-consistent treatment of reflection to date. Particular attention is payed to the coincident evolution of the power-law and reflection, which can be used to determine the accretion geometry. The hard state is found to be distinctly reflection weak, however the ratio of reflection to power-law gradually increases as the source luminosity rises. In contrast the reflection is found dominate the power-law throughout most of the soft state, with increasing supremacy as the source decays. We discuss potential dynamics driving this, favouring inner disc truncation and decreasing coronal height for the hard and soft states respectively. Evolution of the ionisation parameter, power-law slope and high-energy cut-off also agree with this interpretation.
\end{abstract}

%%%%%%%%%%%%%%%%%%%%%%%%%%%%%%%%%%%%%%%%%%%%%%%%%%%%%%%%%%%%%%%%%%%%%%%%%%%%%%%%%%%%%%%%%
%%%%%%%%%%%%%%%%%%%%%%%%%%%%%%%%%%%%%%%%%%%%%%%%%%%%%%%%%%%%%%%%%%%%%%%%%%%%%%%%%%%%%%%%%

\section{Introduction}\label{intro}
The most fundamental classification of black hole X-ray binaries (BHXRBs) indicates two distinct classes: persistent and transient systems. Persistent sources are typically wind-fed by high-mass companions, allowing a consistently high accretion rate and hence modest spectral variability. The transient systems however, which have Roche-lobe filling low-mass companions, exhibit dramatic outbursts spanning as many as 8 orders of magnitude in luminosity, but ultimately spend the majority of their existence in quiescence \citep{Remillard06, Fender12}. Remarkably, almost all transients showcase more or less identical characteristic spectral and temporal evolution in outburst, and understanding the mechanisms driving this has become a key focus in astrophysics. 

At the onset of an outburst the X-ray spectrum is distinctly hard, peaking in power at around 100\,keV, and is well described by power-law ($\Gamma\sim1.6$), hence being commonly known as the low/hard (hereafter `hard') state. In bright hard states an additional soft component is also often observed \citep{DiSalvo01,Miller06,Reis10,Kolehmainen13}, usually attributed to thermal emission with a peak temperature of $\sim0.2$\,keV, whilst there is a high level of aperiodic variability (up to 50\,\%; \citealt{VDK06}). The dominant power-law probably arises from inverse-Compton scattering of `seed' photons, supplied by the accretion disc, in a hot and optically thin `corona' of electrons. Steady radio emission is also observed at GHz frequencies, associated with a steady compact jet \citep{Fender04} and is well correlated with the X-ray emission \citep{Corbel03,Gallo03,Corbel13}. The source spans many decades in X-ray luminosity as it rises up the hard state, but nevertheless shows little spectral evolution.

Above a few \% of L$_{\rm edd}$ systems often (but not always) commence a transition to a softer X-ray state. First the source advances into the hard-intermediate state (HIMS), where now the power-law is steeper ($\Gamma\textgreater2$) and the thermal component has increased in both temperature and contribution (now $\sim50\,\%$), resulting in a softer spectrum (see e.g. \citealt{Hiemstra11}). Progressing through this state the flux continually rises whilst the integrated variability begins to decrease. Beyond the HIMS lies the soft-intermediate state (SIMS) which harbours a similar, albeit slightly softer, spectrum to the HIMS. The timing properties are however distinctly different with a fractional RMS now below 10\,\% \citep{Munoz11}. The flux has also continued to increase, and the SIMS typically marks the most luminous phase of the outburst. As the state transition occurs this also reveals contrasting radio emission, by which the jet declines but also displays dramatic flare events often two or more orders of magnitude more luminous than the steady hard state emission (see e.g. \citealt{M-J12,Brocksopp13}).

After the brief state transition (typically $\sim10$ days) the system enters the high/soft (hereafter `soft') state, dominated by thermal emission from the accretion disc, which now has an effective temperature of $\sim1$\,keV \citep{Dunn11}. The variability lessens further (now $\textless5\,\%$) whilst the Comptonised emission remains steep and relatively weak throughout. Radio emission is now undetected, and is believed to signify quenching of the jet \citep{Russell11}. The soft state typically lasts for many months, during which there is some variation in hardness, and in some cases one or two brief excursions back into the SIMS, and even sometimes the HIMS. Throughout though the source is generally fading in flux, and upon reaching a few \% in Eddington luminosity makes a state transition through the intermediate states and back into the hard state \citep{Maccarone03} where it fades further into quiescence.

The interplay between the respective soft thermal and hard Comptonised emission ultimately defines the two distinct hard and soft states. However, defining the morphology leading to state transitions and separating the two states has proven to be difficult, not least due to an insufficient physical understanding of the corona. Interpreting the role of the jet has also proven to be difficult, even though the X-ray and radio emission correlate well \citep{Corbel03,Gallo03,Corbel13}. Furthermore the base of the jet has been proposed as a source of hard X-rays (e.g. \citealt{Beloborodov99,Markoff05}), heightening the need to understand the connection between the two.

%%%%%%%%%%%%%%%%%%%%%%%%%%%%%%%%%%%%%%%%%%%%%%%%%%%%%%%%%%%%%%%%%%%%%%%%%%%%%%%%%%%%%%%%%

\subsection{The reflection spectrum}

While photons up-scattered in the corona are observed directly as a hard power-law many will also irradiate the disc leading to a number of reprocessing features, collectively known as the \emph{reflection spectrum} (see \citealt{Fabian10} for a recent review). Fluorescent emission, as atoms de-excite after photoelectric absorption, is of highest prominence and interest in the X-ray band, and is dominated by Fe emission as a result of high abundance and fluorescent yield (which varies as the atomic number $Z^4$). Early works by \cite{George91} and \cite{Matt91} performed Monte Carlo calculations of fluorescent emission resulting in estimates of equivalent widths, line strengths and angular dependance for the Fe K line. Of additional importance is electron scattering which dominates above $\sim$10\,keV (photoelectric absorption dominates below this) and is observed as a peak in flux around 20--40\,keV known as the \emph{Compton hump}.

The surface layers of the accretion disc are likely to become ionised by the powerful irradiation arising from the corona, and has lead to many works studying the affect of ionisation upon the reflection spectrum \citep{Ross93,Matt93,Zycki94,Ross99,Nayakshin00,Nayakshin01,Ballantyne01,Ross05}. In particular, \cite{Ross05} represents the grid \textsc{reflionx} which is today the most widely applied reflection model, taking into account the strongest emission lines and self-consistent treatment of the continuum. More recently \cite{Garcia10} introduced \textsc{xillver} (see also \citealt{Garcia11,Garcia13}) which represents a furthering in the treatment of atomic processes, in particular making use of the photo-ionisation code \textsc{xstar}, and is the model applied in this study. The extent of ionisation in the surface layers of the disc is defined by the ionisation parameter $\xi$, which represents the ratio of the illuminating X-ray flux with the gas density \citep{Tarter69,Garcia13}. In BHXRBs lighter elements in the surface layers are expected to be fully stripped by the ionising power of the disc, resulting in a high albedo and values of $\xi$ \citep{Ross93,Zycki94}. This effect also leads to effective Fe emission, since the fully ionised lighter elements are not able to absorb the emitted Fe photons \citep{Matt93}.

As it is inherently dependant upon the geometry of the corona producing the photons and the disc intercepting them, the reflection spectrum presents the opportunity to infer changes in the two components in comparison to their own specific emission.

%%%%%%%%%%%%%%%%%%%%%%%%%%%%%%%%%%%%%%%%%%%%%%%%%%%%%%%%%%%%%%%%%%%%%%%%%%%%%%%%%%%%%%%%%

\subsection{GX 339$-$4 and this study}

GX 339$-$4 is a BHXRB \citep{Hynes03,Munoz08} and one of the most active transient systems, exhibiting numerous outbursts since its discovery \citep{Markert73}, including four complete cycles (with state-changes) in the past twelve years. As a result, GX 339$-$4 is one of the most studied transient systems and over the lifetime of \emph{RXTE}\footnote{Rossi X-ray Timing Explorer} (1995--2012) an extensive archive of data has been amassed, allowing an unparalleled timeline to investigate source variability with the same mission. To this end it has formed the basis of many important works key to our understanding of BHXRBs (see e.g. \citealt{Belloni05,Dunn10,Corbel13} and references therein). Restricting to periods where both the Proportional Counter Array (PCA; \citealt{Jahoda06}) and High Energy X-ray Timing Experiment (HEXTE; \citealt{Rothschild98}) instruments were active this presents three fully sampled outbursts to analyse.

In addition GX 339$-$4 is the best monitored BHXRB in the radio band, allowing a unique insight into the outburst nature of transient systems. In this study we examine how the X-ray reflection evolves throughout these three outbursts, presenting one of the most extensive and detailed studies to date of reflection in BHXRBs. We begin by outlining our data reduction strategy in \S
, followed by details of the model applied to the observations and our automated fitting procedure in \S\ref{model}. We then present the results of the study in \S\ref{analysis} and then outline and discuss our favoured interpretations in \S\ref{discussion}.

%%%%%%%%%%%%%%%%%%%%%%%%%%%%%%%%%%%%%%%%%%%%%%%%%%%%%%%%%%%%%%%%%%%%%%%%%%%%%%%%%%%%%%%%%

\section{Observations and Data Reduction}\label{rxte}

\begin{figure*}
\centering
\epsfig{file=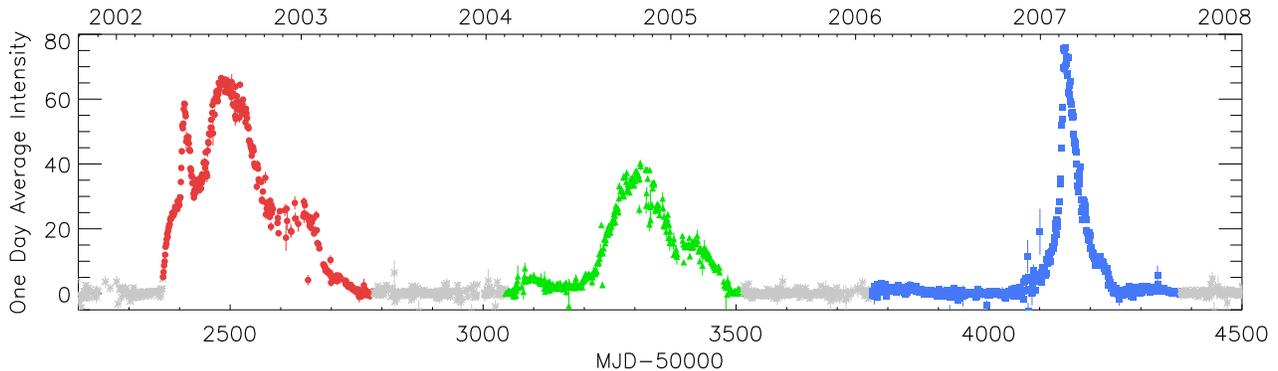,  width=\textwidth}
\caption{The \emph{RXTE} all-sky monitor light-curve of GX 339$-$4 covering the three outbursts analysed in this study. Red (circles), green (triangles) and blue (squares) correspond to the 2002, 2004 and 2007 outbursts respectively.}
\label{ASM}
\end{figure*}

%%%%%%%%%%%%%%%%%%%%%%%%%%%%%%%%%%%%%%%%%%%%%%%%%%%%%%%%%%%%%%%%%%%%%%%%%%%%%%%%%%%%%%%%%

We performed spectral analysis using data from the PCA and the HEXTE onboard \emph{RXTE}. The data were reduced using HEASOFT software package v6.13 following the standard steps described in the (RXTE) data reduction cookbook\footnote{http://heasarc.gsfc.nasa.gov/docs/xte/data\_analysis.html}. We extracted PCA spectra from the top layer of the Proportional Counter Unit (PCU) 2 which is the best calibrated detector out of the five PCUs, although we added a systematic uncertainty of 0.5\,\% to all spectral channels to account for any calibration uncertainties. We produced the associated response matrix and modelled the background to create background spectra.

For HEXTE, we produced a response matrix and applied the necessary dead-time correction. The HEXTE background is measured throughout the observation by alternating between the source and background fields every 32s. The data from the background regions were then merged. When possible we used data from both detector A and B to extract source and background spectra. However, from 2005 December, due to problems in the rocking motion of Cluster A, we extracted spectra from Cluster B only. HEXTE channels were grouped by four. The variability analysis (i.e. the RMS-Intensity diagrams) was performed using event modes or a combination of single-bit modes depending on the case.

\begin{figure}
\centering
\epsfig{file=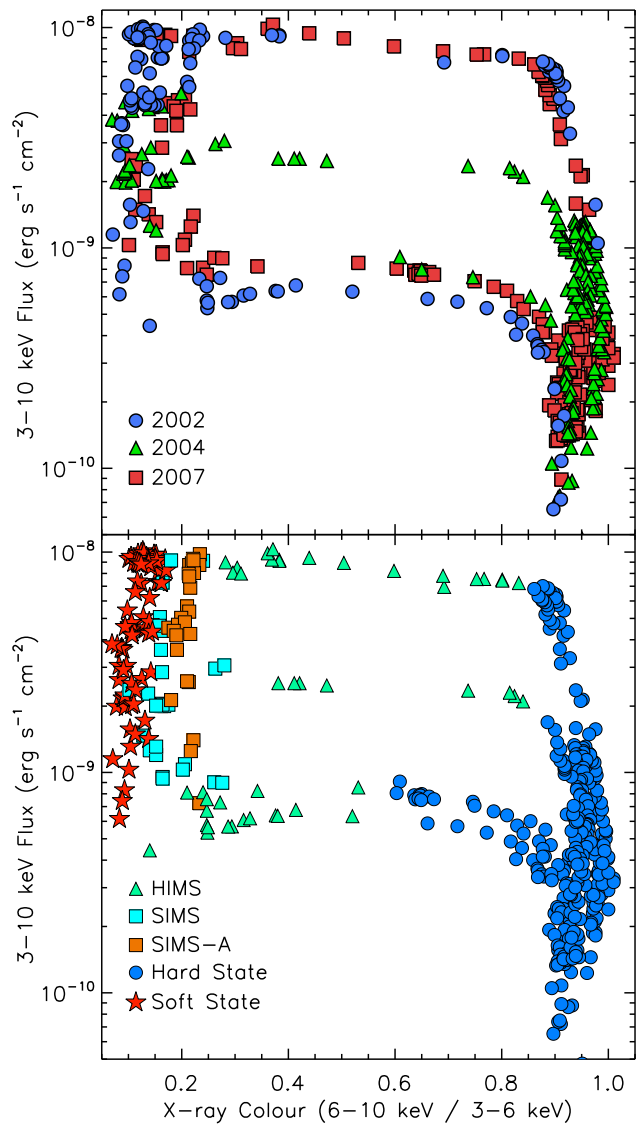, width=0.5\textwidth}
\caption{Hardness-intensity diagrams of all the observations analysed in this study. Top and bottom figures correspond to symbols determined by outburst (Table \ref{observations}) or state (\S\ref{states}).}
\label{HID}
\end{figure}
\begin{table}
\centering
\begin{tabular}{llll}\hline
Outburst		& Start (MJD)	& End (MJD)	& Total Observations (Exposure)		\\\hline	
2002			& 52367		& 52784		& 133 (300 ks)						\\
2004			& 53044		& 53514		& 202 (465 ks)						\\
2007			& 53769		& 54379		& 193 (357 ks)						\\\hline
Total			&			&			& 528 (1122 ks)					\\\hline	
\end{tabular}
\caption{The dates and number of observations used in this study after our selection criteria outlined in \S\ref{procedure} has been applied. The threshold for a dataset to be analysed is defined in \S\ref{model}, such that observations at low flux ({\it i.e.} the periods of quiescence) and of poor quality are ignored. We also list the respective amount of observations and the subsequent total exposure time per outburst.}
\label{observations}
\end{table}

%%%%%%%%%%%%%%%%%%%%%%%%%%%%%%%%%%%%%%%%%%%%%%%%%%%%%%%%%%%%%%%%%%%%%%%%%%%%%%%%%%%%%%%%%
%%%%%%%%%%%%%%%%%%%%%%%%%%%%%%%%%%%%%%%%%%%%%%%%%%%%%%%%%%%%%%%%%%%%%%%%%%%%%%%%%%%%%%%%%

\section{Fitting method}\label{model}

There are two significant reasons we have used archival \emph{RXTE} observations for this study. Firstly, the PCA (3--50\,keV) and HEXTE (25--200\,keV) instruments provide an outstanding spectral bandpass. Although the Fe K line is the prominent feature of X-ray reflection, it is ultimately just a fraction of the overall signal. When looking for subtle effects, such as those due to strong gravity, the line profile is a very revealing signature. However the poor spectral resolution of the PCA ($\sim$ $1$\,keV at $6$\,keV) renders it insufficient to attempt precise constraints on parameters such as black hole spin. However, the large bandpass and effective area allows us to study a much larger fraction of reflection spectrum, including important signatures such as the Compton hump and Fe edge, which characterise emission over a much larger range. As a result of this, \emph{RXTE} offers an alternative and complementary view to high-resolution missions such as \emph{XMM-Newton}, which are restricted to the 0.3--10\,keV band.

This leads on to the second significant rationale, which is the unprecedented number of observations that \emph{RXTE} offers. In a recent study of the hard state of GX 339$-$4 limited to high-resolution \emph{XMM-Newton} and \emph{Suzaku} spectra, \cite{Plant13} were restricted to just four observations, whereas the sampling of the hard state in this investigation amounts to nearly 350 observations. In total this study makes use of more than 500 observations, totalling over $1.1$\,Ms of PCA exposure time, allowing one of the most detailed investigations to date of how the reflection spectrum is evolving throughout an outburst (Table \ref{observations} and Figure \ref{HID}). In particular, such high cadence enables a scrupulous analysis of state transitions, which are typically completed within one week.

%%%%%%%%%%%%%%%%%%%%%%%%%%%%%%%%%%%%%%%%%%%%%%%%%%%%%%%%%%%%%%%%%%%%%%%%%%%%%%%%%%%%%%%%%

\subsection{Our adopted model}

Many attempts have been made to quantify how X-ray reflection evolves in BHXRBs, but often they have employed phenomenological models focusing upon the Fe K region. \cite{Dunn08} fitted the Fe K$\alpha$ emission with a single \textsc{gaussian} line fixed at 6.4\,keV, while \cite{Rossi05} applied relativistic broadened emission (\textsc{laor}; \citealt{Laor91}) with a smeared edge (\textsc{smedge}; \citealt{Ebisawa94}). Both offer a useful diagnostic of the reflection, however they only account for a small fraction of the full reflection signal. Furthermore while the latter may offer a more physical interpretation of the Fe K$\alpha$ line, the two components are still not physically linked. Ultimately, to fully understand how the reflection component is evolving one must apply a model where all the signatures are physically linked and accounted for. Motivated by this, \cite{Reis13} recently applied the blurred reflection model \textsc{kdblur*reflionx} \citep{Laor91,Ross05} to an outburst of XTE J1650-500. Although this presented a significant improvement to previous works, it only focuses upon one outburst of one source. Here we apply an angle-dependent version of the reflection model \textsc{xillver} \citep{Garcia10,Garcia11}, known as \textsc{xillver-a} \citep{Garcia13}, to a much larger sample (Table \ref{observations}). This model offers a significant improvement in the treatment of atomic processes over the more widely used \textsc{reflionx}, and in addition includes larger grid for the photon index and ionisation parameters (see \citealt{Garcia13} for a comparison). It should be noted that \textsc{reflionx} represents an angle-averaged solution, hence neglecting the important effect of inclination in the observed reflection spectrum \citep[see e.g.][]{Magdziarz95,Matt96,Garcia14}.

We account for interstellar absorption using the model \textsc{phabs} with a N$_{\rm H}$ fixed at $0.5\times 10^{22}$\,cm$^{-2}$, given that the hydrogen column density towards GX 339$-$4 is well resolved to be within the range (0.4--0.6)$\times10^{22}$\,cm$^{-2}$ \citep{Kong00}. Since the low-energy cut-off of the PCA is at $\sim 3$\,keV the ability to constrain the column freely is severely reduced and we also note that the effect of this moderate column above $3$\,keV is small.

The remaining continuum is characterised by a combination of a multi-colour blackbody (\textsc{diskbb}; \citealt{Mitsuda84}) and Comptonised seed photons (\textsc{cutoffpl}). Recently it has been shown by \cite{Munoz13} that relativistic effects can strongly affect how we interpret the thermal emission. However such impact is fairly weak at the low-moderate inclination GX 339$-$4 is expected to have (\citealt{Munoz13} classify GX 339$-$4 as a `low inclination' source), hence \textsc{diskbb} should describe the thermal accretion disc emission sufficiently well. We model the Comptonised emission as a simple power-law with an exponential cut-off. While more physical Comptonisation models exist, they ultimately bring with them further complexity in fitting the data at hand, and is beyond the scope of this study focusing upon reflection. 

Another important consideration is the consistency between the continuum we resolve and the spectral energy distribution assumed in the reflection model. The illuminating spectrum used in \textsc{xillver-a} assumes a power-law with a high-energy cut-off at $300$\,keV, making \textsc{cutoffpl} the best Comptonisation model to maintain consistency and allowing an accurate determination of relative flux levels. When a smaller cut-off is required we switch to a solution of \textsc{xillver-a} including a variable cut-off for the illuminating spectrum, linking the cut-off in the Comptonisation and reflection models. \textsc{xillver-a} also includes a low-energy cut-off at 0.1\,keV to prevent the input spectrum being unphysically over-populated with low-energy photons. It should be noted though that the reflection spectrum will also be influenced by photons emitted in the disc \citep{Ross93,Ross07}, and this is not currently accounted for by \textsc{xillver-a}, nor is it in any other publicly available reflection model\footnote{Note: The model \textsc{rfxconv} \citep{Done06,Kolehmainen11} can accept any input continuum, including that of a disc, which we discuss in \S\ref{accuracy}}. Finally, we also apply the convolution model \textsc{relconv} \citep{Dauser10} to account for any relativistic effects. Thus our base model applied to each observation is:

\textsc{phabs(diskbb + cutoffpl + relconv$\ast$xillver-a)}

\renewcommand{\arraystretch}{1.2}
\begin{table}
\centering
\begin{tabular}{lllll}
\hline
Model						& Parameter				& Value		& Min		& Max		\\\hline	
\textsc{phabs}					& N$_{\rm H}$ ($10^{22}$)	& 0.5			&			& 			\\
\textsc{diskbb}					& T$_{\rm in}$				& 0.5			& 0.1			& 2			\\
							& N$_{\rm D}$				& 1000		& 0			& 1e+10		\\
\textsc{cutoffpl}					& $\Gamma$				& 2			& 1.2			& 3.4			\\
							& E$_{\rm c}$				& 50			& 10			& 300		\\
\textsc{relconv}					& $\theta$ ($^{\circ}$)		& 45			& 			& 			\\
							& r$_{\rm in}$ (r$_{\rm g}$)	& 10			& 6			& 1000		\\
							& r$_{\rm out}$ (r$_{\rm g}$)	& 1000		& 			& 			\\
							& $\alpha$				& 3			& 			& 			\\
							& $a$					& 0			& 			& 			\\
\textsc{xillver-a}				& $\log(\xi)$				& 2.5			& 1 			& 4.5			\\
							& $\theta$ ($^{\circ}$)		& 45			& 			& 			\\\hline							
\end{tabular}
\caption{A list of input parameter values and hard limits for our base model applied to each dataset in this study. Parameters listed without limits indicate that it is fixed at that stated value. The input photon index in the \textsc{xillver-a} model is linked to that of \textsc{cutoffpl}. We also assume solar abundances and a single power-law emissivity index (R$^{-\alpha}$) for the reflection.}
\label{inputpar}
\end{table}

%%%%%%%%%%%%%%%%%%%%%%%%%%%%%%%%%%%%%%%%%%%%%%%%%%%%%%%%%%%%%%%%%%%%%%%%%%%%%%%%%%%%%%%%%

\subsection{The fitting procedure}\label{procedure}

By definition transient BHXRBs spend the majority of their lifetime in a quiescent state. GX 339$-$4 is somewhat an exception to this being one most active of all the transient systems, exhibiting four complete outbursts in the past twelve years. Nevertheless, observations during periods of inactivity, or not reaching sufficient statistics through low count rates and short exposures, will require removal. For an observation to be considered we require a PCA pointing with at least 1000 background subtracted counts. We ignore all data below 3\,keV (channels $\le4$) and above 50\,keV (channels $\ge83$), however the regions $\ge20$\,keV and $\ge35$\,keV must have a minimum of 100 and 50 background subtracted counts respectively to be included. Furthermore these must amount to at least 10\% of the total counts registered in that region.

Signatures of the reflection spectrum at hard X-ray energies, such as the Compton hump, offer a large signal in addition to the Fe K line. Additionally, the reflection remains strong at high energies allowing greater constraint. This hence underlines why we take such an approach to the upper-energy threshold of the PCA and a similar criteria is therefore applied to the HEXTE to ensure adequate data quality. We require that the combined HEXTE A and B cluster units, if both available (see \S\ref{rxte}), have a minimum of 1000 background subtracted counts. If this is not the case then the observation is ignored regardless of the merit of the PCA. We fix the lower energy limit to be 25\,keV and allow the upper bound to be 200\,keV, however this is truncated to 100\,keV should there be less than 100 background subtracted counts above this threshold.

In the hard and some hard-intermediate states, fitting the disc was problematic due to the low-energy cut-off of the PCA. Thermal emission from the disc in the hard state has been uncovered in many sources using instruments extending to softer energies (\textless1\,keV; see e.g. \citealt{DiSalvo01,McClintock01,Kolehmainen13}), however the signal above 3\,keV is not sufficient to constrain the \textsc{diskbb} parameters. For this reason if T$_{\rm in} \textless 0.1$\,keV then \textsc{diskbb} is removed to ensure the analysis is reliable. In such cases the model is likely to be fitting some slight curvature in the Comptonised continuum or having a negligible effect. We also remove discs with T$_{\rm in} \textgreater 2$\,keV since temperatures are not expected to reach such levels, especially for a low inclination source (see e.g. \citealt{Dunn11,Munoz13}), and hence suggests a erroneous fit. Additionally the normalisation value must exceed 50 to flag any severe degeneracy with the disc temperature, and should it be below this value then the disc model is removed. We note that normalisation is of the order $10^3$ in the soft state, where the disc is expected to be at the innermost stable circular orbit (ISCO), and should hence represent the lowest values in the outburst. These criteria do not affect the soft and soft-intermediate state observations, all of which retain the disc model within this criteria.

\begin{figure}
\centering
\epsfig{file=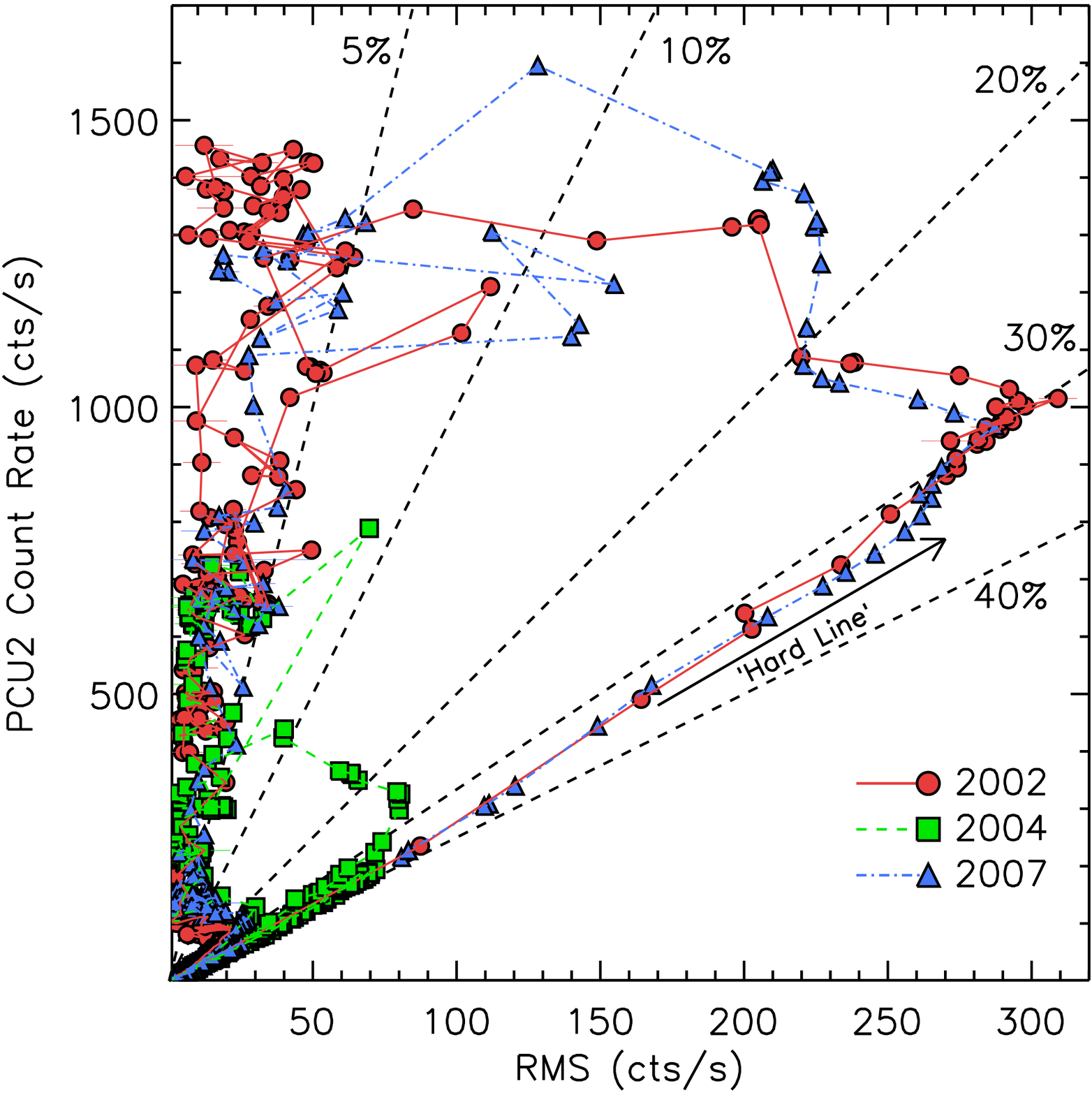, width=0.5\textwidth}\\
\epsfig{file=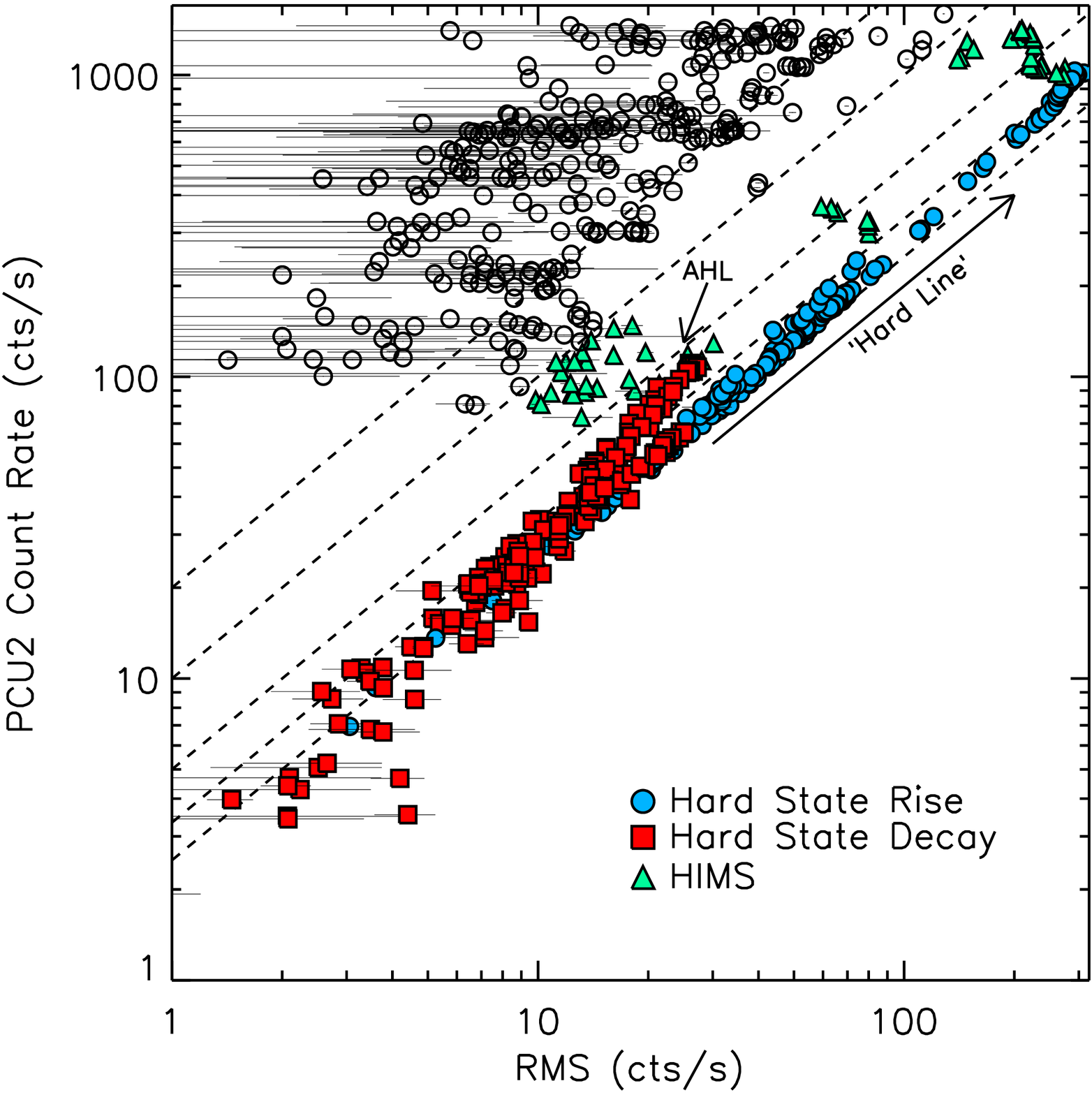, width=0.5\textwidth}
\caption{RMS-intensity diagrams of the three outbursts of GX 339$-$4 featured in this study (data taken from \citealt{Munoz11}). The black dashed lines represent fractional RMS levels. Top: Symbols are colour-coded by outburst and plotted linear-linear to view the entire outburst. Bottom: The same data with a log-log scale to focus upon the rise (blue) and decay (red) of the hard state. Open circles represent observations not in the hard state or HIMS. The `adjacent hard line' (AHL) signifies the return to the hard state.}
\label{RID}
\end{figure}

One of the free parameters of the reflection model \textsc{xillver-a} is the photon index of the source illuminating the ionised slab (i.e. the accretion disc), of which the tabulated range is 1.2--3.4. We link this parameter to the \textsc{cutoffpl} photon index to keep the two components self-consistent using the tabulated range as the hard limits. Previous studies of BHXRBs suggest this range should be sufficient, however it is another motivation for us using \textsc{xillver-a} instead of \textsc{reflionx}, which offers a smaller grid of 1.4--3.3. The high energy cut-off (E$_{\rm c}$) is allowed to be free within the range $10$ to $300$\,keV, however if the parameter reaches \textless $15$\,keV or \textgreater $250$\,keV the cut-off is fixed at $300$\,keV for the remainder of the analysis. This prevents any erroneous fits and maintains consistency with the assumed illuminating spectrum in the standard version of \textsc{xillver-a}. Only in the bright states of the hard state is a smaller high energy cut-off required, hence during this period we switch to a larger grid of \textsc{xillver-a} which includes the high energy cut-off of the illuminating spectrum as a fitted parameter. In this case the parameter is linked between the \textsc{cutoffpl} and \textsc{xillver-a} models to maintain consistency. We note that the values and evolution of the high energy cut-off are consistent with that found by Motta et al. (2009; see Figure \ref{cutoff} and \S\ref{PL} later in this paper).

The other key parameters in \textsc{xillver-a} are the ionisation parameter, $\xi=4\pi F/n_{\rm e}$ (where $F$ is the total illuminating flux in the 1-1000 Ry energy band and $n_{\rm e}$ the electron number density; \citealt{Tarter69,Garcia13}), the iron abundance (which we assume to be solar), and the inclination angle of the source. The ionisation parameter is fitted within the range $\log$(1.0--4.5) and a fixed inclination of 45$^{\circ}$ is applied. To date there is no precise measurement of the inclination of GX 339$-$4, however a value lower than $\sim40^{\circ}$ appears unfeasible given that it will result in BH mass of $\textgreater20$M$_{\sun}$ (assuming the constraints reported in \citealt{Hynes03} and \citealt{Munoz08}). The value of 45$^{\circ}$ is consistent with the findings of \cite{Plant13} through the reflection method and the lack of evidence for a accretion disc wind, which are visible only in high inclination sources ($i\,\gtrsim\,70^{\circ}$; \citealt{Ponti12}). Also, we note that dips and eclipses have never been detected from GX 339$-$4.

The relativistic effects are modelled using \textsc{relconv}, for which the emissivity index $\alpha$, defined as R$^{-\alpha}$ where R is the disc radius, is fixed to be 3. While the spectral resolution of the PCA ($\sim$ $1$\,keV at $6$\,keV) allows moderately broad lines to be detected, the more subtle effects of the emissivity profile cannot be resolved, thus the Newtonian value of 3 is reasonable assumption for this study. We do however allow the inner radius to be fitted freely from the ISCO (6\,r$_{\rm g}$) up to the largest tabulated value of 1000\,r$_{\rm g}$. The outer radius of the disc is fixed to be 1000\,r$_{\rm g}$ and the black hole is assumed to have zero spin. The effect of the latter is rather trivial given the resolution of the PCA and is nevertheless not a significant concern given that the relativistic effects are not the main focus of this study. Finally the inclination is fixed at 45$^{\circ}$ to be consistent with the reflection model.

Comparing relative flux levels can be hampered by the chosen energy band, hence all fluxes are extrapolated using \textsc{cflux} and fitted in the 0.1--1000\,keV band unless otherwise stated. As an example, Comptonised photons reprocessed in the cool disc may be down-scattered to energies below the \emph{RXTE} bandpass, hence extrapolating the fit ensures that the truest extent of the spectrum is being measured. The lower limit is chosen to ensure that there is not an unphysical over-population of photons in the cut-off power-law model, which is particularly a problem when the photon index is large ($\Gamma\,\textgreater\,2$). The illuminating spectrum in the reflection model also assumes a low-energy cut-off of 0.1\,keV to tackle this problem, hence we are therefore keeping the two models consistent by imposing this limit.

To deal with such a vast dataset we fit each observation through an automated routine, therefore Table \ref{inputpar} lists the input parameter values applied to our adopted model. Throughout the investigation we use XSPEC version 12.8.0 and all quoted errors are at the 90\% confidence level unless otherwise stated. Throughout the study we apply fiducial values of and 10\,$M_{\sun}$ and 8\,kpc for the black hole mass and distance to GX 339$-$4 respectively.

%%%%%%%%%%%%%%%%%%%%%%%%%%%%%%%%%%%%%%%%%%%%%%%%%%%%%%%%%%%%%%%%%%%%%%%%%%%%%%%%%%%%%%%%%

\subsection{Defining spectral states}\label{states}

In order to separate each observation into spectral states we use a combination of timing and spectral characteristics. While the hard and soft states are easily discernible via their spectra in the PCA bandpass, the intermediate periods are qualitatively alike. \cite{Munoz11} studied the relation between the RMS amplitude of the variability and flux (Figure \ref{RID}), uncovering marked changes in the transition between states in the three outbursts of GX 339$-$4 used in this study. The rising hard state follows a distinctive `hard track' of increasing RMS and flux, with a fraction RMS of 30--40\,\%. Transition into the HIMS occurs as the RMS lessens whilst the flux continues to rise. A further significant decrease in RMS marks the SIMS, which we define as the region $\textless$10\,\% fractional RMS.

\begin{figure}
\centering
\epsfig{file=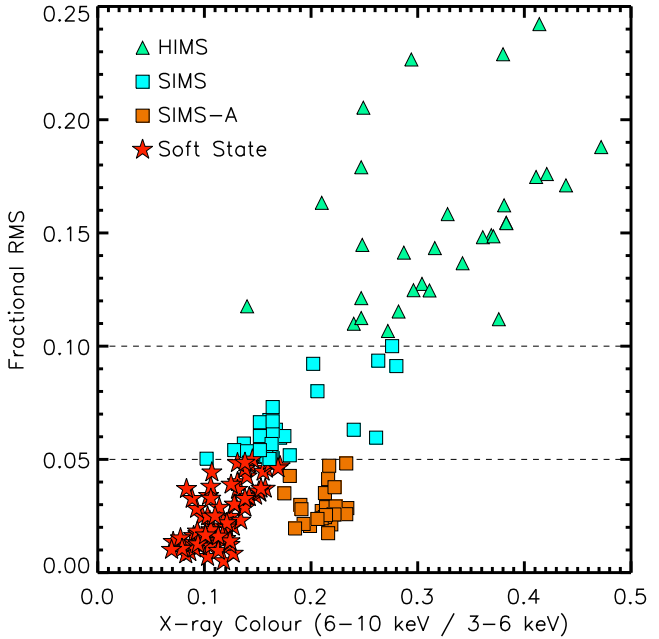, width=0.5\textwidth}
\caption{Fractional RMS versus spectral hardness displaying how the intermediate and soft spectral states are determined (\S\ref{states}). The SIMS-A are a cluster of points with low variability (fractional RMS $\textless$5\,\%) and a relatively high spectral hardness ($\textgreater$0.175\,\%), which we propose as a distinct spectral state (see \S\ref{states} and \S\ref{SIMS-A}). The hard state is determined by the RMS-intensity relation (Figure \ref{RID}).}
\label{RMS_hard}
\end{figure}

The soft-state is characterised again by a sudden decrease in RMS, and defined here by a fractional RMS of $\textless$5\,\%. However there exists a group of harder observations (X-ray colour $\textgreater$0.175; Figure \ref{RMS_hard}) at low variability. We note as well that these points lie typically at a slightly higher fractional RMS than the least variable soft-state points. The state classification for these observations is not straight forward since they share properties associated to both SIMS and soft state \citep[see e.g.][]{Belloni05,Belloni11}. Nevertheless, we define these points as SIMS-A since many show type-A quasi-periodic oscillations (QPOs), which is in contrast to the normal SIMS (5\,$\textless$\,RMS\,$\textless$\,10) where type-B oscillations are observed \citep{Motta11}. Interestingly the SIMS-A lie off the standard luminosity-temperature relation established by the soft state (\S\ref{L-T}). We do note however that the 5\,\% line separating the soft and SIMS should be taken with caution since, for example, fast transitions occurs as this stage of the outburst, leading to hybrid observations \citep{Munoz11}. To determine the HIMS to hard state transition we use 'adjacent hard line' (AHL) described in \cite{Munoz11}, whereby the decay phase of the hard state displays a clear track in the RMS-intensity diagram. This is indicated by the red circles in the lower plot of Figure \ref{RID}. The transition is not so clear if one attempts to determine it by X-ray colour (Figure \ref{HID}), and thus demonstrates the importance of timing characteristics in determining the state of the system.

%%%%%%%%%%%%%%%%%%%%%%%%%%%%%%%%%%%%%%%%%%%%%%%%%%%%%%%%%%%%%%%%%%%%%%%%%%%%%%%%%%%%%%%%%
%%%%%%%%%%%%%%%%%%%%%%%%%%%%%%%%%%%%%%%%%%%%%%%%%%%%%%%%%%%%%%%%%%%%%%%%%%%%%%%%%%%%%%%%%

\section{Analysis and results}\label{analysis}

\subsection{The Comptonised emission}\label{PL}

\begin{figure*}
\centering
\epsfig{file=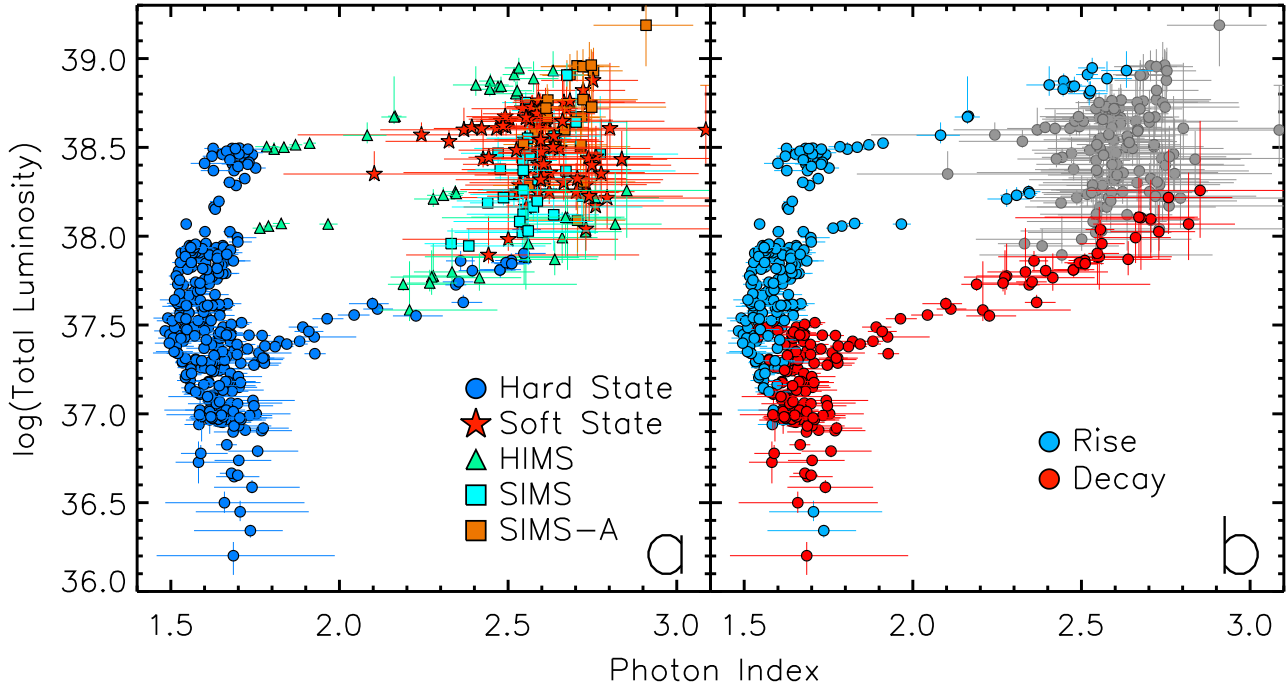,  width=\textwidth}
\caption{The total source luminosity plotted against the power-law photon index $\Gamma$. Left: Points coloured by their respective state. Right: The same diagram, however the soft state, SIMS and SIMS-A observations are now grey, with the HIMS and hard state separated according to the rise or decay phase of the outburst. Observations are plotted randomly to avoid any visual bias. The y-axis is plotted in units of erg s$^{\rm-1}$.}
\label{gamma}
\end{figure*}

In Figure \ref{gamma} we display how the photon index $\Gamma$ evolves with the total unabsorbed luminosity. The shape is remarkably similar to the HID (Figure \ref{HID}) and indicates well how the hardness of the X-ray spectrum is not solely due to the influence of the thermal emission from the accretion disc. The blue points in Figure \ref{gamma}a mark the hard state whereby $\Gamma$ mainly lies in a narrow range of 1.5--1.7, and appears to remain rather constant despite the hard state spanning over two orders of magnitude in luminosity. However, Figure \ref{gamma}b separates the periods of rise (blue) and decay (red) and displays quite clearly that above $L_{\rm X}\sim10^{37.5}$ $\Gamma$ tends to become softer as the source rises up the hard track, consistent with an increase in seed photons to cool the corona (see \citealt{Done07} and references therein). Furthermore as the source reaches the hard state in decay the photon index is clearly softer than the rise at that luminosity, appearing to retain a softer slope towards quiescence (although see also \citealt{Stiele11}).

During the state transition the photon index undergoes very distinct softening, eventually reaching $\Gamma\sim2.5$. In fact the bulk of the evolution appears to take place in the HIMS, rather than the SIMS, which itself is consistent with the values recorded in the soft state. The soft state shows some scatter (2.4--2.9), but is ultimately dominated by large confidence intervals due to the diminished signal in the hard band. The SIMS-A are also typically steeper than the SIMS (see also Table \ref{avpar}). The decay phase (Figure \ref{gamma}b) displays clear hardening as the source makes its way through the soft-hard transition. In Figure \ref{gamma} we plot against the total luminosity calculated between 0.1 and 1000\,keV which accounts for the non-negligible disc flux below 3\,keV, therefore displaying the source decay through the transition not so apparent in the 3--10\,keV HID (Figure \ref{HID}). This then serves to exhibit clearly how the source hardens monotonically with luminosity towards the hard state.

Throughout our investigation, we model the Comptonised emission with a cut-off power-law fixed at 300\,keV, in order to remain consistent with the assumed illuminating spectrum in the reflection model \citep{Garcia13}. However, as part of our routine, we initially allow the high energy cut-off to be a free parameter as it has been shown to be significantly lower in the brighter phases of the hard state and subsequent transition \citep{Motta09}. We allow the cut-off to be free within the range 10 to 300\,keV, however should it converge on a value lower than 15\,keV or higher than 250\,keV respectively we deem the fit to be erroneous and fix the parameter to be 300\,keV.

\begin{figure}
\centering
\epsfig{file=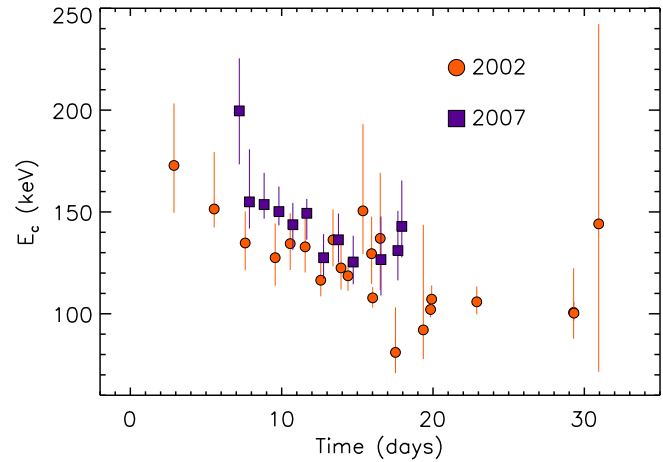,  width=0.5\textwidth}
\caption{Evolution of the high energy cut-off with time. Zero marks the first observation where the cut-off is detected (MJD 52372 and 54122 for the 2002 and 2007 outbursts respectively). No cut-off was resolved in the 2004 outburst observations.}
\label{cutoff}
\end{figure}

In Figure \ref{cutoff} we plot the time-evolution of the high energy cut-off in the 2002 and 2007 outbursts. No cut-off was resolved in any of the 2004 observations, however the transition luminosity in this outburst was lower than any of the 2002 and 2007 observations which did. Time-zero in Figure \ref{cutoff} marks the first observation where the cut-off was determined freely and displays a gradual decrease in energy as the source continues its rise through the hard state. All observations intermediate to the respective first and last resolved observation in Figure \ref{cutoff} were able to fit the parameter freely. Furthermore, for both outbursts, the final three observations resolving the cut-off took place in the HIMS, confirming this trend continues into the state transition. The high energy cut-off is thought to represent the temperature of the electrons in the corona, hence meaning the power-law emission symbolises thermal Comptonisation. The softening of the photon index and decreasing high energy cut-off are both consistent with increasing amounts of seed photons cooling the corona.

The trend we find is very consistent with that found by \cite{Motta09} using the same datasets, however they resolve the cut-off over a longer period than us all the way into the soft state. In this study we are more focused on the reflection, hence to maintain a reasonable timescale for our routine we do not apply such a stringent and detailed criteria for detecting the cut-off, hence we refer the interested reader to the work of \cite{Motta09} for an more in depth study of the cut-off.

%%%%%%%%%%%%%%%%%%%%%%%%%%%%%%%%%%%%%%%%%%%%%%%%%%%%%%%%%%%%%%%%%%%%%%%%%%%%%%%%%

\subsection{The reflection fraction ($RF$)}\label{RF}
\begin{figure*}
\centering
\epsfig{file=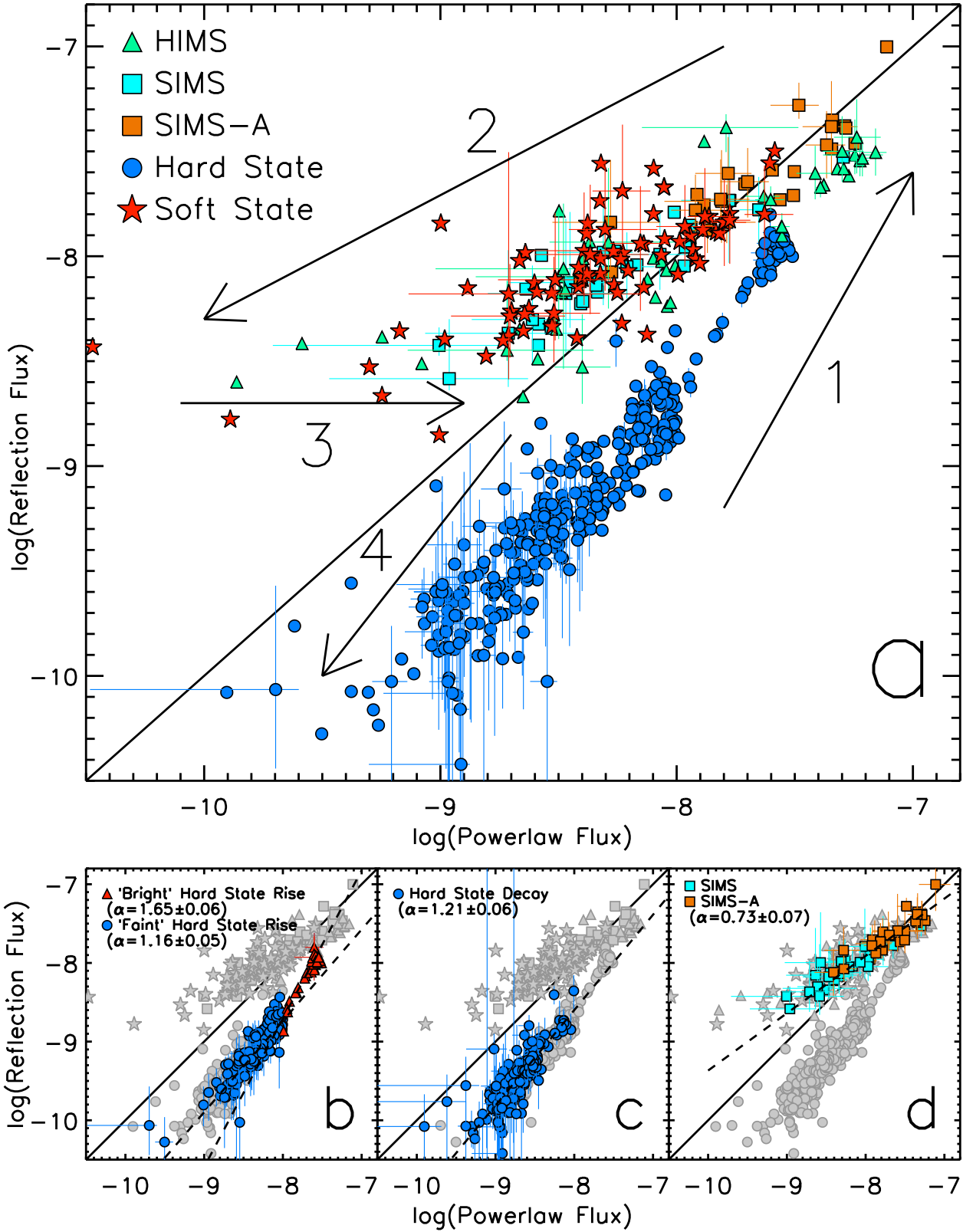, width=0.95\textwidth}
\caption{This figure displays how the power-law and reflection components evolve in outburst via their respective fluxes. Both scales are identical and the solid line indicates one-to-one quantities. Top: Numbered arrows illustrate how the source progresses throughout the plot in outburst. For clarity we only plot confidence limits for the 2007 outburst in this figure. Bottom: These plots focus on specific states of the outburst with dotted lines representing the best-fit to the highlighted observations, such that $\log_{10} S_{\rm PL}=\alpha\log_{10} S_{\rm Ref}+c$. Left: The rising phase of the hard state, separated into observations with a power-law flux less than (blue circles) or greater than (red triangles) $10^{-8}$\,erg s$^{\rm-1}$cm$^{\rm-2}$ respectively. Middle: The decaying phase of the outburst (towards quiescence). Right: SIMS and SIMS-A points, fitted together. Errors on the fitted slope are stated at the $1\sigma$ confidence level. Each axis is displayed in units of erg s$^{\rm-1}$cm$^{\rm-2}$.}
\label{PL_R}
\end{figure*}

The illumination of the disc by Compton up-scattered photons gives rise to the reflection spectrum, thus one would expect the two components to follow a relatively linear relationship. As an example, increasing the total of Comptonised photons by a factor of two should lead to the same doubling of the power-law flux we observe, and that irradiates the disc, assuming the geometry remains constant. Therefore in such a case the reflected emission will double as well. Ultimately the situation is not that simple as the surface layers of the disc will also respond to the change, for example by becoming more ionised which may in turn affect the albedo (but see \S\ref{Xi_1}). Nevertheless it serves as a useful example of how the power-law and reflected emission should evolve somewhat identically \textit{if} the accretion geometry does not vary. Alternatively, if the two components do not evolve in a one-to-one manner then there is a considerable chance that the geometry is changing. Thus observing how these two components evolve together presents a very powerful diagnostic to reveal the underlying accretion dynamics in outburst.

We find that indeed the power-law and reflection are very well correlated, forming strong positive correlations throughout each spectral state (Table \ref{pFlux}). Here we define the reflection fraction (\emph{RF}) as the ratio of the reflection to the Comptonised flux, such that $\emph{RF}=1$ corresponds to an equal quota of both components. Figure \ref{PL_R} details the evolution of the two elements throughout the three outbursts studied, with the five principle states colour-coded. The hard state (blue circles) lies exclusively below the solid line indicating $\emph{RF}=1$, thus the hard state Comptonised flux always dominates the reflection. The fitted slope to the rising hard track is $\sim1.5$, however it indicates two distinct regions hence we instead separate the rising hard state into observations above and below a power-law flux of $10^{-8}$ erg s$^{\rm-1}$cm$^{\rm-2}$ (Figure \ref{PL_R}b). At lower luminosities (`faint hard state') the slope is $1.16\pm{0.05}$, thus signifying a gradually increasing reflection fraction as the total source flux rises. Here the mean reflection fraction is $\sim$0.2. The track of the `bright hard state' is much steeper ($1.65\pm{0.06}$) marking a clear furthering in how the rate of the reflection fraction is increasing, which reaches values up to 0.6. We note that whilst this is coincident with the period where the high energy cut-off is fitted below $300$\,keV (\S\ref{PL}), this should not be behind this effect as the reflection spectrum includes the same cut-off in the illuminating spectrum it is calculated from. In addition we fit the decay phase of the hard state as the source returns to quiescence (Figure \ref{PL_R}c), for which the slope is consistent with the `faint' regime of the rise ($1.16\pm{0.05}$ vs $1.21\pm{0.06}$), thus suggesting that the underlying changes responsible for the evolution in the reflection fraction are consistent whether the source is at the onset or end of the outburst. To summarise, we find that there is a substantial increase in the reflection fraction as the overall source flux rises, although it remains less than one throughout the hard state. In addition, the $RF$ correlates better with the reflection component, rather than the Comptonised one, suggesting changes in the former are more behind the evolution of the $RF$ (Table \ref{pFlux}). 

\renewcommand{\arraystretch}{1.2}
\begin{table*}
\centering
\begin{tabular}{llllllllllllll}
\multicolumn{12}{c}{Spearman's Rank Correlations For Various Fitted Parameters}\\\hline
				&					& \multicolumn{2}{l}{Hard State}	& \multicolumn{2}{l}{HIMS}				& \multicolumn{2}{l}{SIMS}		& \multicolumn{2}{l}{SIMS-A}		& \multicolumn{2}{l}{Soft State}	\\\hline
Parameter 1		& Parameter 2			& $\rho$	& $p$-value			& $\rho$		& $p$-value				& $\rho$	& $p$-value			& $\rho$	& $p$-value			& $\rho$	& $p$-value			\\\hline
$S_{\rm powerlaw}$	& $S_{\rm reflection}$	& 0.938	& 0					& 0.871		& 3.88$\times10^{-15}$		& 0.916 	& 4.90$\times10^{-13}$	& 0.900	&4.17$\times10^{-10}$	& 0.735	& 1.31$\times10^{-14}$	\\	
$S_{\rm powerlaw}$	& $S_{\rm disc}$		& ...		& ...					& 0.710$^{\star}$	& 1.82$\times10^{-6}$$^{\star}$	& 0.694	& 1.52$\times10^{-5}$ 	& 0.839	& 8.40$\times10^{-8}$	& 0.474  	& 1.04$\times10^{-5}$	\\
$S_{\rm reflection}$	& $S_{\rm disc}$		& ...		& ...					& 0.731$^{\star}$	& 6.02$\times10^{-7}$$^{\star}$	& 0.601	& 3.52$\times10^{-4}$	& 0.653 	& 3.02$\times10^{-4}$	& 0.438	& 5.43$\times10^{-5}$	\\
$RF$			& $S_{\rm disc}$		& ...		& ...					& $-$0.528$^{\star}$	& 1.12$\times10^{-3}$$^{\star}$	& $-$0.701 	& 1.13$\times10^{-5}$	& $-$0.825	& 2.17$\times10^{-7}$	& $-$0.361	& 1.07$\times10^{-3}$	\\
$RF$			& $S_{\rm powerlaw}$	& 0.375	& 5.73$\times10^{-13}$	& $-$0.835		& 5.64$\times10^{-13}$		& $-$0.855	& 8.94$\times10^{-10}$ 	& $-$0.688	& 1.04$\times10^{-4}$	& $-$0.765  	& 2.30$\times10^{-16}$	\\
$RF$			& $S_{\rm reflection}$	& 0.637	& 1.01$\times10^{-40}$	& $-$0.498		& 4.35$\times10^{-4}$		& $-$0.621	& 1.96$\times10^{-4}$ 	& $-$0.366	& 6.58$\times10^{-2}$	& $-$0.190  	& 9.32$\times10^{-2}$	\\
\hline	
\end{tabular}
\caption{ A list of SpearmanÕs rank correlations ($\rho$) and the the corresponding false correlation probability ($p$-value) calculated for the flux determined in each model in the five principle spectral states. The parameter $S$ refers to the respective flux component.
\newline$^{\star}$Calculated using only observations requiring a disc model.}
\label{pFlux}
\end{table*}

The HIMS points (green triangles) are situated almost entirely at larger $\emph{RF}$ than those of the hard state. Furthermore they are typically close to $\emph{RF}=1$. There are however two clusters of points from the hard-soft branch due to the lower flux of the transition in the 2004 outburst. The soft-hard branch (S$_{\rm reflection}\sim10^{-8.5}$ erg s$^{\rm-1}$cm$^{\rm-2}$) leads gradually but directly onto the hard track suggesting that the process in decay is progressive rather than sudden. The SIMS(-A) points lie at larger $\emph{RF}$ to the HIMS in the hard-soft transition (those at higher fluxes), continuing the general rise observed so far throughout the states (see also Table \ref{avpar}). The soft state itself is almost entirely consistent with a $\emph{RF}\textgreater1$, meaning that the reflection is now dominant over the Comptonised component. There is however a lot of scatter, and large confidence limits due to the customary reduced signal in the hard band, thus the slope of the soft state decay can only be fitted when the 2002 outburst is excluded, but nevertheless this confirms that the $RF$ increases as the source decays ($\alpha=0.49\pm0.17$). The soft state also tends to show significant variations in variability and spectral hardness throughout its decay (see Figures \ref{HID} and \ref{RID}) and thus makes several forays into the SIMS and SIMS-A between state transitions. The best-fit to the SIMS and SIMS-A observations yields a slope of $0.73\pm{0.07}$ (Figure \ref{PL_R}d) representing an furthering of the domination by the reflection component over the Comptonised emission as the source decays, consistent with the evolution of the reflection and power-law in the soft state. Throughout the soft and soft-intermediate states the $RF$ correlates best with the Comptonised component, suggesting that it is mainly responsible for the increasing $RF$ in decay (Table \ref{pFlux}). The $RF$ is a proxy for the EW of the Fe line, hence this agrees with the soft-state X-ray Baldwin effect previously reported in GX 339$-$4 (\citealt{Dunn08}; see also \citealt{Baldwin77}).

%%%%%%%%%%%%%%%%%%%%%%%%%%%%%%%%%%%%%%%%%%%%%%%%%%%%%%%%%%%%%%%%%%%%%%%%%%%%%%%%%%%%%%%%%

\subsection{Contrasting emission in spectral states}\label{spectra}

\begin{figure*}
\centering
\epsfig{file=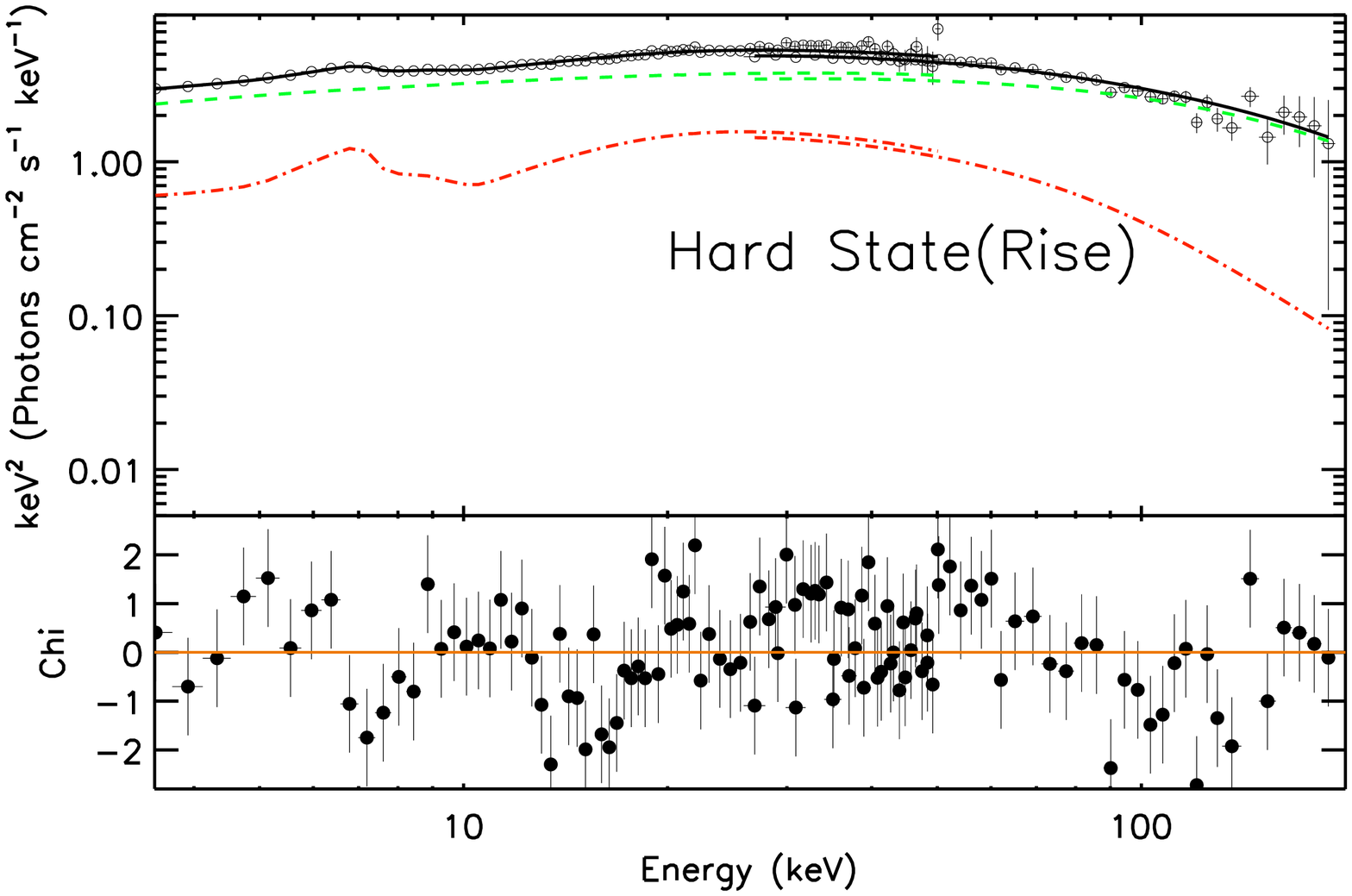, width=0.48\textwidth}
\epsfig{file=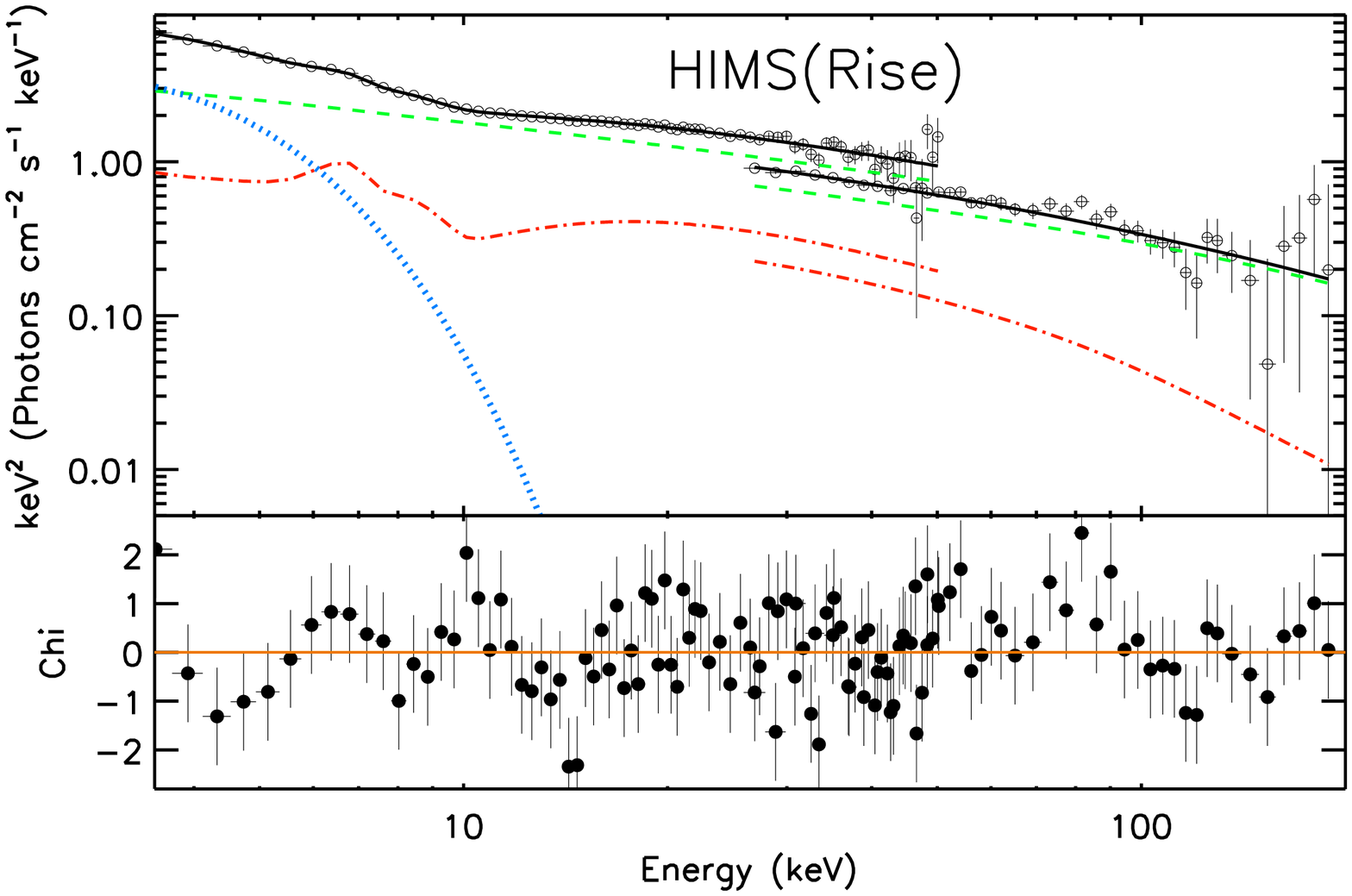, width=0.48\textwidth}\\
\epsfig{file=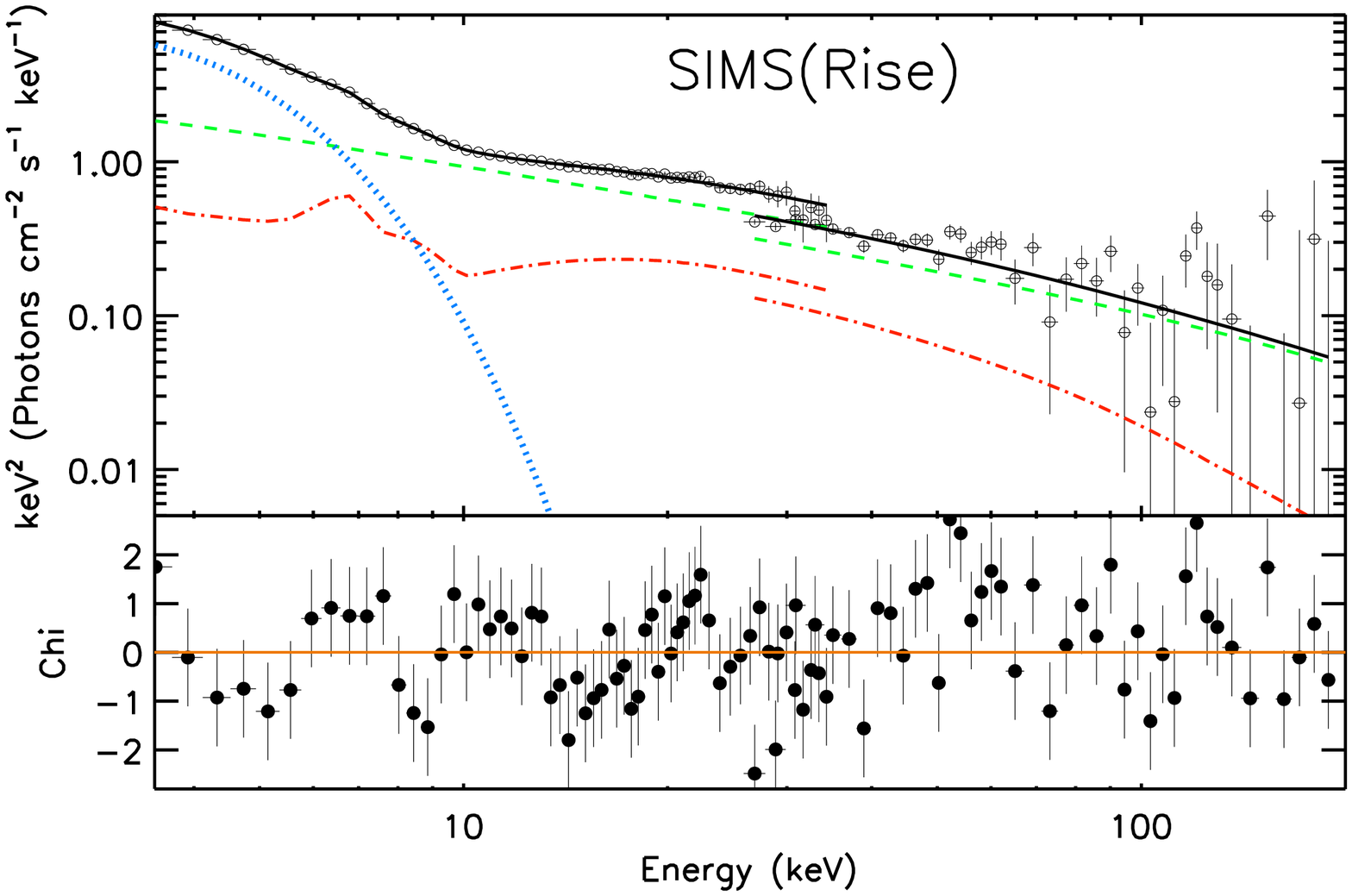, width=0.48\textwidth}
\epsfig{file=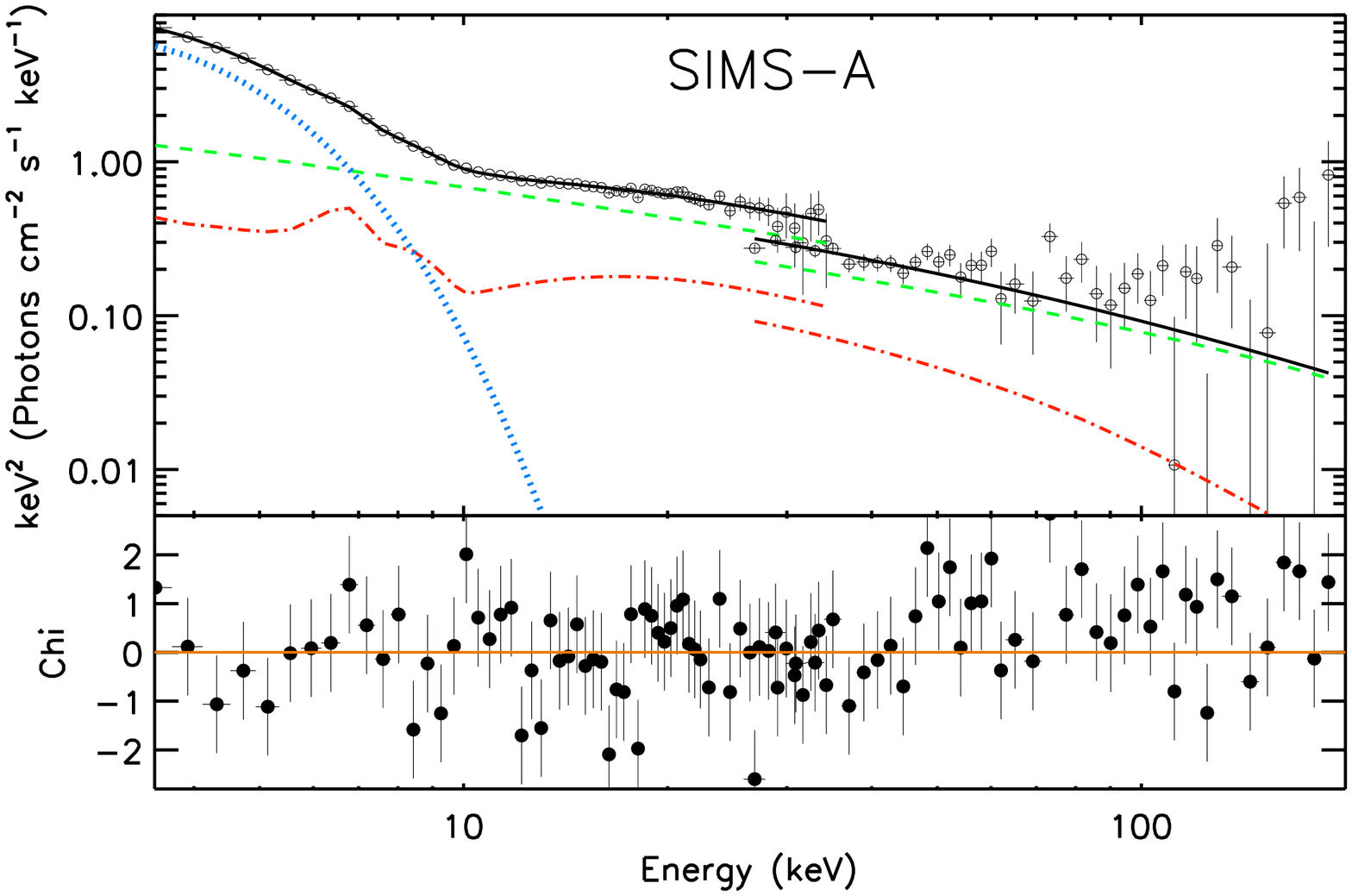, width=0.48\textwidth}\\
\epsfig{file=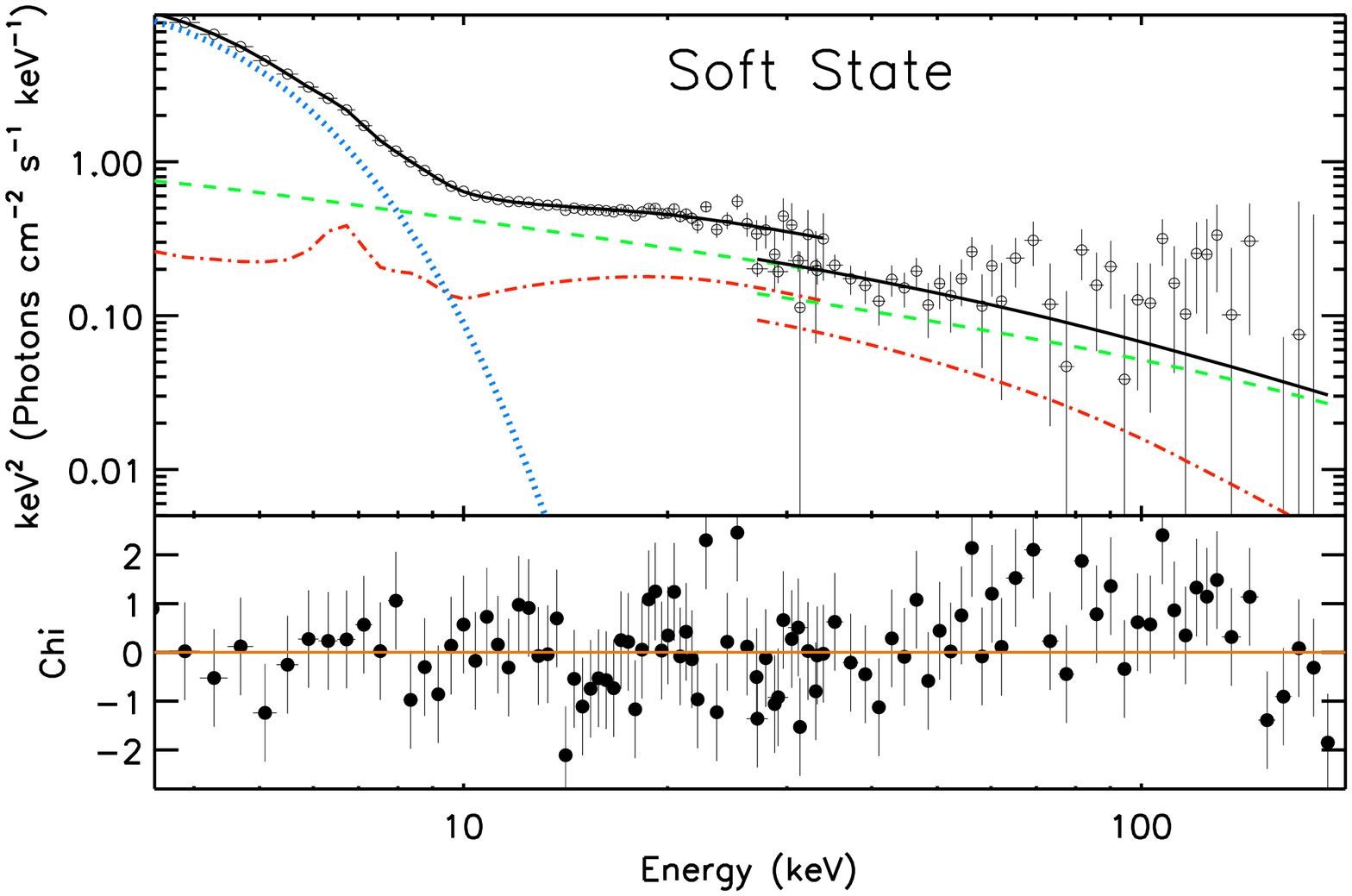, width=0.48\textwidth}
\epsfig{file=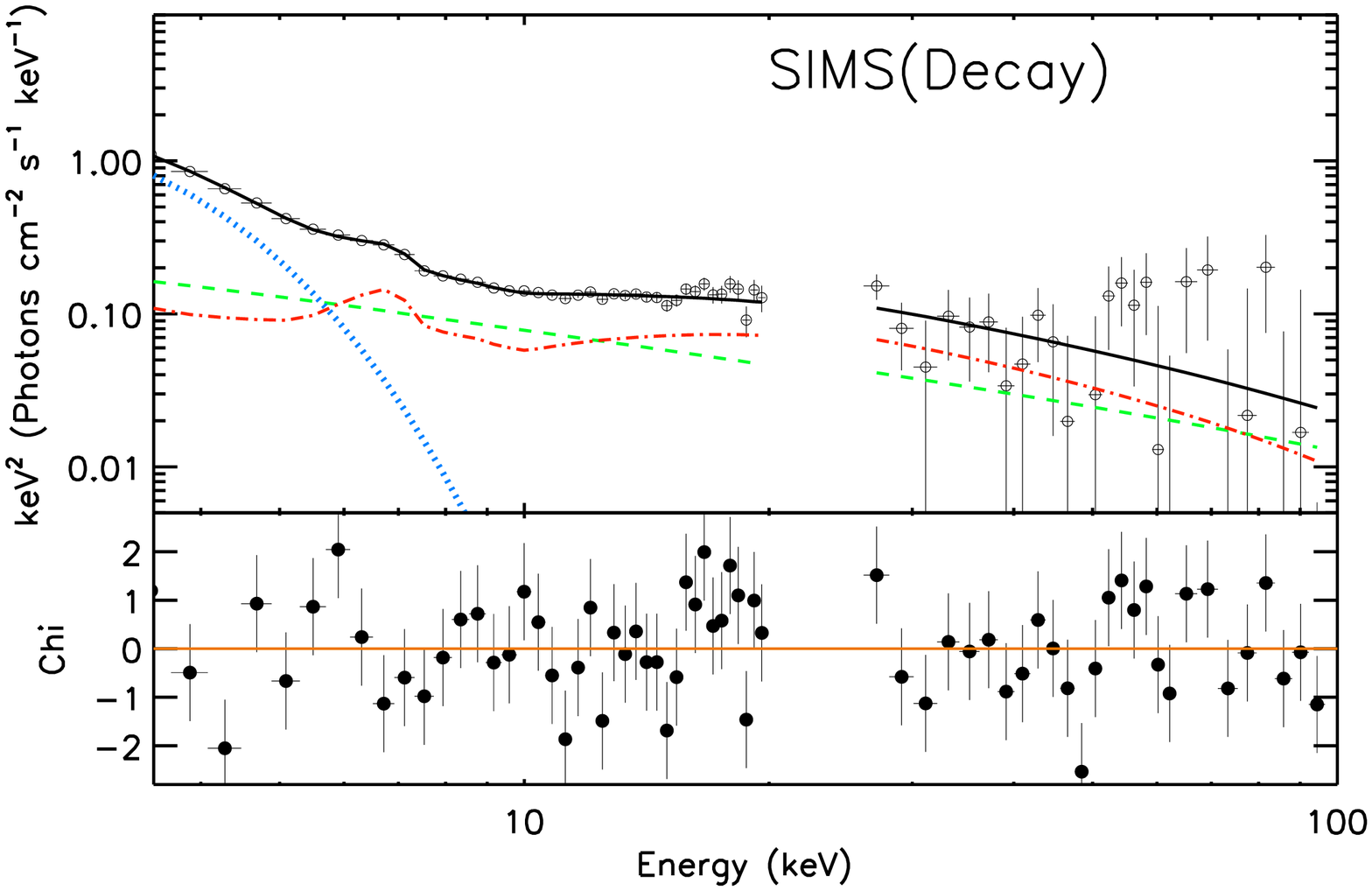, width=0.48\textwidth}\\
\epsfig{file=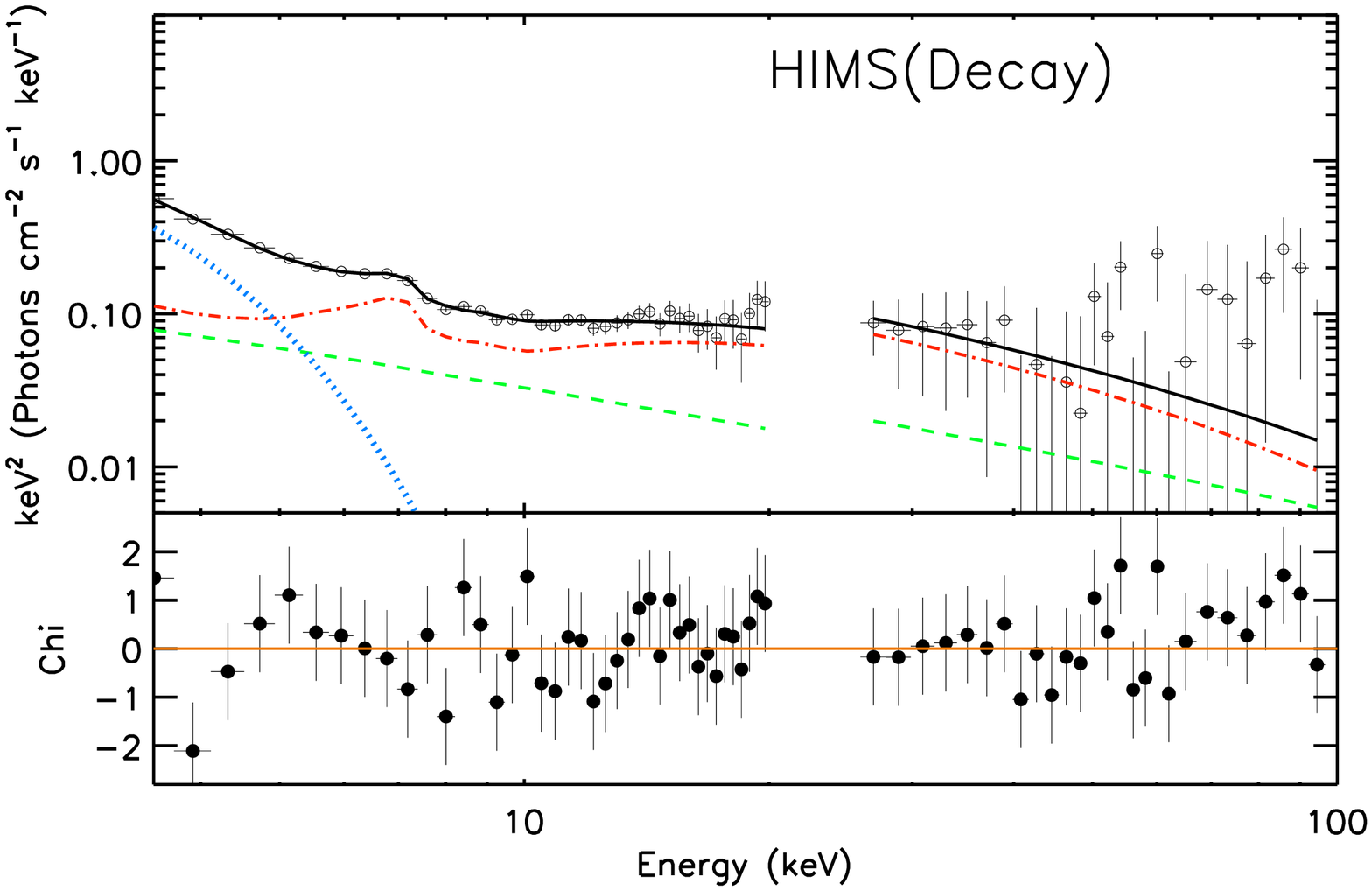, width=0.48\textwidth}
\epsfig{file=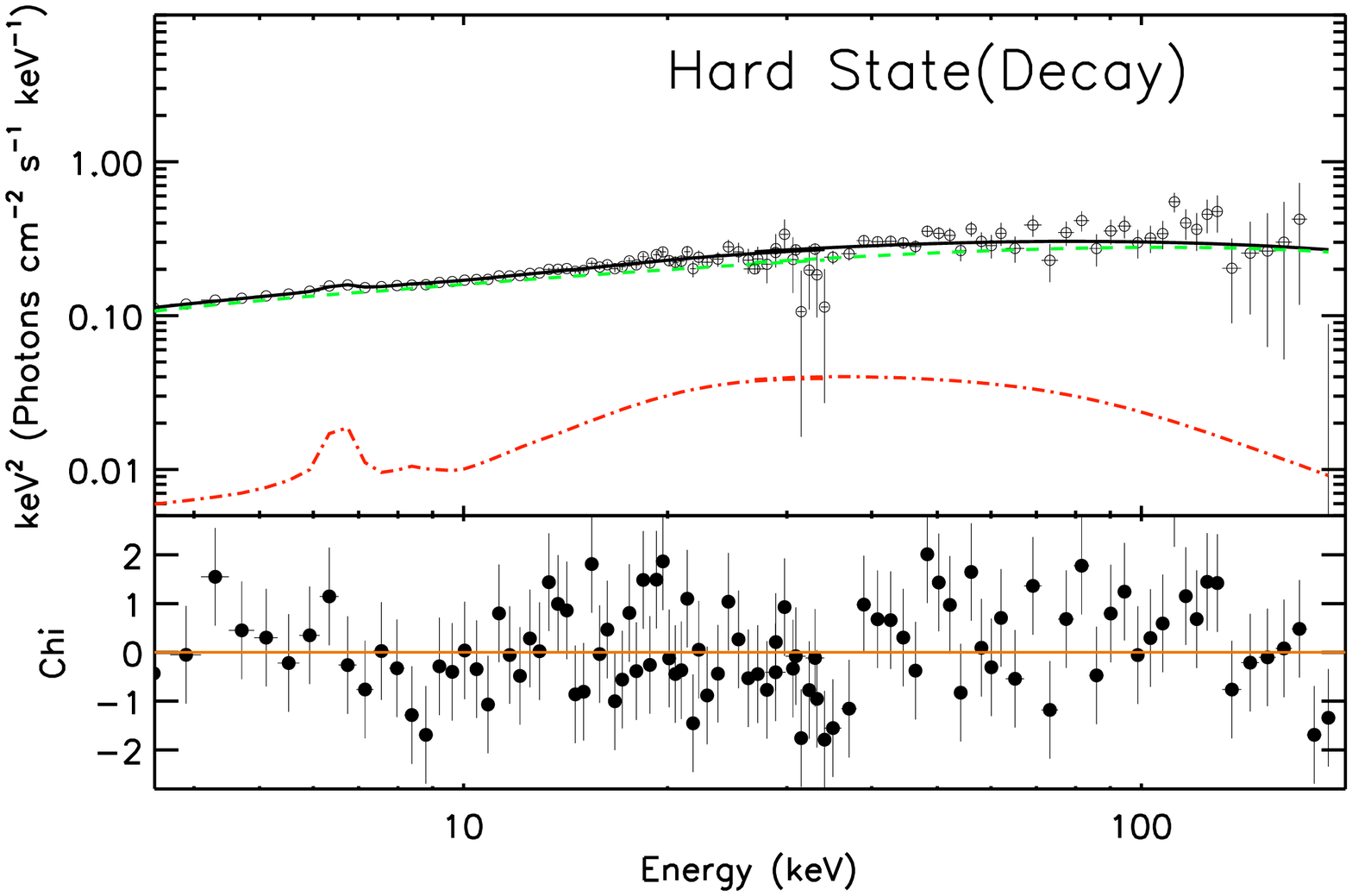, width=0.48\textwidth}
\caption{Unfolded spectra (top) and data residuals (bottom) for the eight states listed in Table \ref{avpar}. The disc, cut-off power-law, reflection and total models are displayed by blue (dotted), green (dashed), red (dot-dash) and black (solid) lines respectively. The data is also plotted as black open circles. Each vertical scale is equivalent to best exemplify how the spectrum evolves in outburst. Areas where the model overlap and appear to disagree are due to the calibration constant between the PCA and HEXTE instruments.}
\label{fig_states}
\end{figure*}

In Figure \ref{fig_states} we display how the X-ray spectrum evolves throughout each spectral state, whilst in Table \ref{avpar} we calculate the mean parameter value accordingly. In the rising hard state a disc model is never required, whilst the Comptonised emission is hard and clearly the dominant source of emission. The Fe line profile is often narrow and the high-energy cutoff is clearly visible in the plotted spectrum. Transition to the HIMS marks increases in the power-law slope, ionisation parameter and reflection fraction. Thermal emission from the disc is now clearly present, however it remains the weaker of the three spectral components and only dominates below $\sim3$\,keV.

Into the SIMS there is a distinct softening, in part due to a further increase of $\Gamma$. The disc is now much stronger, dominating up to $\sim$6--7\,keV, and is the largest source of emission. A broad Fe line is now regularly observed and the ionisation stage is considerably higher than in the hard state. In addition reflection and Comptonised emission are roughly equal. The SIMS-A is only subtly different from the SIMS, the most marked difference being a slightly softer $\Gamma$ (see also Figure \ref{gamma}). Typically the total flux is larger, in fact the SIMS-A represents the peak luminosity for two of the three outbursts in this study. Table \ref{avpar} suggests that this increase is due to greater flux from the power-law and reflection rather than the disc (see also Figure \ref{PL_R}d to compare power-law and reflection flux between the SIMS and SIMS-A).

\renewcommand{\arraystretch}{1.2}
\begin{table*}
\centering
\begin{tabular}{llllllllll}
\hline\multicolumn{9}{c}{Mean Parameter Value Per Spectral State}\\\hline
Parameter				& HS(Rise)		& HIMS(Rise)		& SIMS(Rise)		& SIMS-A			&HSS			& SIMS(Decay)		& HIMS(Decay)			& HS(Decay)			\\\hline	
T$_{\rm in}$				& ...				& $0.86\pm{0.09}^{\star}$	& $0.76\pm{0.07}$	& $0.82\pm{0.08}$	& $0.79\pm{0.09}$	& $0.67\pm{0.04}$	& $0.65\pm{0.08}^{\star}$		& ...					\\
N$_{\rm D}$				& ...				& $958\pm{474}^{\star}$	& $2430\pm{645}$	& $1870\pm{219}$	& $3249\pm{194}$	& $2159\pm{678}$	& $1101\pm{641}^{\star}$	& ...					\\
$\Gamma$				& $1.60\pm{0.06}$	& $2.22\pm{0.30}$	& $2.57\pm{0.09}$	& $2.71\pm{0.12}$	& $2.60\pm{0.14}$	& $2.55\pm{0.10}$	& $2.53\pm{0.22}$		& $1.76\pm{0.24}$		\\
r$_{\rm in}$ (r$_{\rm g}$)$^{\dagger}$		& $255\pm{358}$	& $101\pm{259}$	& $44\pm{17}$		& $47\pm{32}$		& $177\pm{298}$	& $50\pm{44}$		& $28\pm{12}$			& $416\pm{449}$		\\
$\log(\xi)$					& $2.75\pm{0.52}$	& $3.44\pm{0.23}$	& $3.40\pm{0.04}$ 	& $3.48\pm{0.25}$	& $3.26\pm{0.20}$	& $3.42\pm{0.09}$	& $3.40\pm{0.20}$		& $2.33\pm{0.69}$		\\\hline
\multicolumn{9}{c}{Mean Model Luminosity ($10^{37}$\,erg s$^{-1}$)}\\\hline
Power-law				& $7.23\pm{6.20}$	& $24.4\pm{15.8}$	& $8.50\pm{8.39}$	& $21.6\pm{14.4}$	& $4.98\pm{4.34}$	& $2.63\pm{1.99}$	& $2.04\pm{1.07}$		& $1.79\pm{1.34}$		\\
Reflection					& $2.13\pm{2.78}$	& $15.6\pm{8.15}$	& $9.22\pm{4.58}$	& $21.5\pm{14.3}$	& $7.93\pm{4.72}$	& $4.61\pm{1.81}$	& $4.82\pm{2.97}$		& $0.39\pm{0.45}$		\\	
Disc						& ...				& $8.84\pm{4.88}^{\star}$	& $13.6\pm{5.73}$	& $14.5\pm{5.40}$	& $21.4\pm{4.72}$	& $7.28\pm{3.64}$	& $2.53\pm{1.06}^{\star}$		& ...					\\
$RF$					& $0.23\pm{0.10}$	& $0.81\pm{0.56}$	& $1.42\pm{0.66}$	& $1.16\pm{0.48}$	& $2.43\pm{2.22}$ 	& $2.12\pm{0.75}$	& $4.00\pm{4.70}$		& $0.20\pm{0.13}$		\\\hline
\end{tabular}
\caption{The mean value for a number of interesting parameters calculated for each spectral state. The errors represent the standard deviation and are thus only meant as an indication of accuracy.
\newline The parameter N$_{\rm D}$ is the normalisation of the \textsc{diskbb} model.
\newline$^{\star}$Calculated using only observations requiring a disc model.
\newline$^{\ddagger}$Calculated from the reflection component.}
\label{avpar}
\end{table*}

\begin{figure}
\centering
\epsfig{file=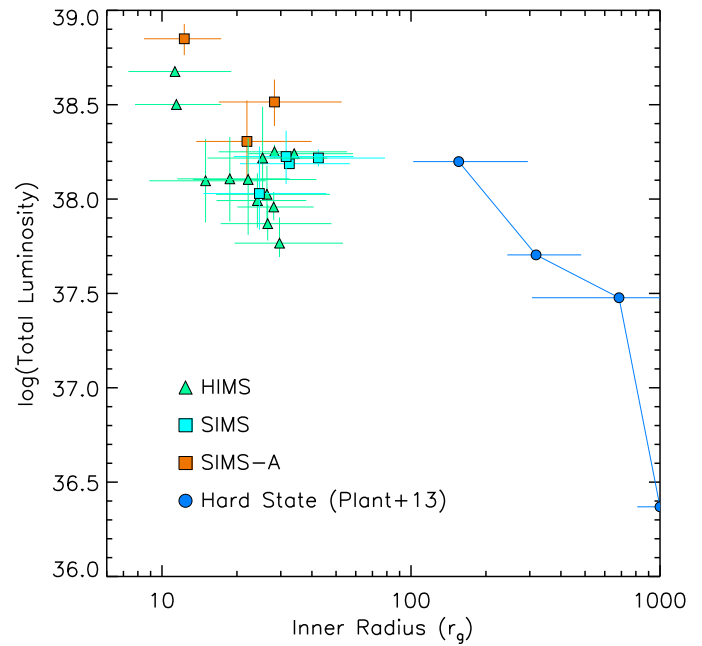, width=0.5\textwidth}
\caption{The fitted inner radius parameter from the reflection component plotted against the total model luminosity for the intermediate states. We restrict the sample to ensure the parameter was well constrained: both the upper and lower error of the inner radius must less than the value of the fitted inner radius, which itself must not be at the hard limits of 6\,$r_{\rm g}$ or 1000\,$r_{\rm g}$. For comparison we plot the hard state results of Plant et al. (2013) who applied the same reflection model and a very similar inclination value (42.1$^{\circ}$ vs 45$^{\circ}$) to high-resolution data. The HIMS, SIMS and SIMS-A appear to represent a stable inner radius, while in the hard state the disc is truncated.}
\label{rin}
\end{figure}

The soft state represents the relative peak of the thermal disc emission which now dominates the spectrum up to 10\,keV. The reflection flux is significantly larger than that of the power-law, which itself decreases substantially as the source decays. Regardless of this, the disc inner radius is often poorly resolved, which we discuss in the next section. As the source decays through the intermediate states the spectral characteristics are similar to the rise, albeit at a lower luminosity. The cooler disc does however contribute a much larger portion to the overall flux, and the reflection fraction is typically larger. The ionisation stage remains relatively unchanged, and the inner radius is once again well resolved. Finally the decay phase of the hard state is much like the rise, described by a hard spectrum and a return to a low reflection fraction and ionisation.

%%%%%%%%%%%%%%%%%%%%%%%%%%%%%%%%%%%%%%%%%%%%%%%%%%%%%%%%%%%%%%%%%%%%%%%%%%%%%%%%%%%%%%%%%

\subsection{The disc inner radius and the ISCO}

\subsubsection{The inner radius from reflection}

The disc inner radius is poorly determined (a larger error) in the hard state, which is likely due to the reduced signal of Fe line in low reflection fractions. At this stage of the outburst the line is also expected to be narrow (\citealt{Plant13,Kolehmainen13}; but see also \citealt{Reis10}), hence making limits on the relativistic broadening difficult with the energy resolution of the PCA ($\sim$ $1$\,keV at $6$\,keV). 

The inner radius is also poorly resolved in the soft state, which is likely to be due to a combination of factors. Firstly, whilst the soft state has a high reflection fraction this only describes the ratio of the reflected and Comptonised emission, and hence does not necessarily mean the reflection is dominating the spectrum. In Figure \ref{fig_states} is can be seen that even at 6 keV the disc flux is significantly larger than the reflection flux, and simply saturates the signal of the Fe emission. Furthermore, since the disc is likely to be at the ISCO in the soft state the Fe profile will be very broad and hence smeared into the continuum. In addition, since the three spectral components are at similar flux levels in the soft state, subtle changes can easily broaden or narrow a weak Fe profile, resulting in the large scatter of inner radii recorded. Recently \cite{Kolehmainen11} showed that even with the spectral-resolution of \emph{XMM-Newton} the width of the Fe can be strongly affected by the continuum, and urge caution in measuring the inner disc radius in the soft state. Our ability to resolve the inner radius is certainly hampered by the resolution of \emph{RXTE} and susceptible to model degeneracy. Despite this, we note that the reflection spectrum is still well constrained because of the large bandpass above 7\,keV.

There is much less scatter in the inner radius fitted during the HIMS and SIMS(-A), both in rise and decay, suggesting that the parameter is being well resolved. Furthermore the mean value is similar throughout (30--50\,$r_{\rm g}$; Table \ref{avpar}) which would indicate that there is no significant change in the disc through these states. The inner radius is known to be very degenerate with the inclination, which is unknown for GX 339$-$4, thus we can not rule out that the inner radius is smaller and represents the ISCO. However, the assumed inclination is very similar to that fitted by \cite{Plant13} in a study of the hard state of GX 339$-$4 with \emph{XMM-Newton} and \emph{Suzaku} allowing a direct comparison. In Figure \ref{rin} we plot the fitted inner radius against source luminosity using only well constrained values (limits less than the parameter value) and include the values found in \cite{Plant13}. Clearly the HIMS and SIMS(-A) are associated with a inner disc closer to the black hole during the intermediate states, while this figure again indicates the consistent radii recorded throughout this stage. To compare, the inner radius derived from the normalisation of the \textsc{diskbb} model is $4.19\pm{0.13}\,r_{\rm g}$ in the soft state for the same inclination (\S\ref{norm_d}).

\subsubsection{The luminosity-temperature relation}\label{L-T}

The soft thermal component is associated with a geometrically thin, optically thick accretion disc \citep{Shakura73}. Under the assumption that the energy released by the disc thermalises, and is emitted locally, the component can be fit as a sum of blackbody spectra from each radius, R. Towards smaller radii the emitting area decreases, whilst the luminosity increases, thus leading to an increase in the characteristic temperature $T_{\rm eff}$(R). The peak temperature occurs close to the ISCO. 

The model we use to fit the thermal component is \textsc{diskbb} \citep{Mitsuda84}, which is parameterised by the apparent inner radius of the disc $r_{\rm in}$ and the peak temperature $T_{\rm in}$. The bolometric disc luminosity is thus defined as $L_{\rm disc} = 4\pi r_{\rm in}^2\sigma T_{\rm in}^4$, where $\sigma$ is the Stefan-Boltzmann constant \citep{Done07}. However, there are a number of corrections required to determine the effective peak temperature, two being the boundary condition \citep{Gierlinski99} and relativistic corrections \citep{Zhang97}, although for a stable inner radius both should remain constant. The spectrum will also be modified by the opacity of this disc, which requires a colour-correction factor $f_{\rm col}$ of 1.6--2.0 \citep{Shimura95,Merloni00,Davis05}. Assuming there is no variation in $f_{\rm col}$  we can therefore test whether the inner radius $r_{\rm in}$ is constant by assessing whether the expected $L_{\rm disc} \propto T_{\rm in}^4$ relation is satisfied.

\begin{figure*}
\centering
\epsfig{file=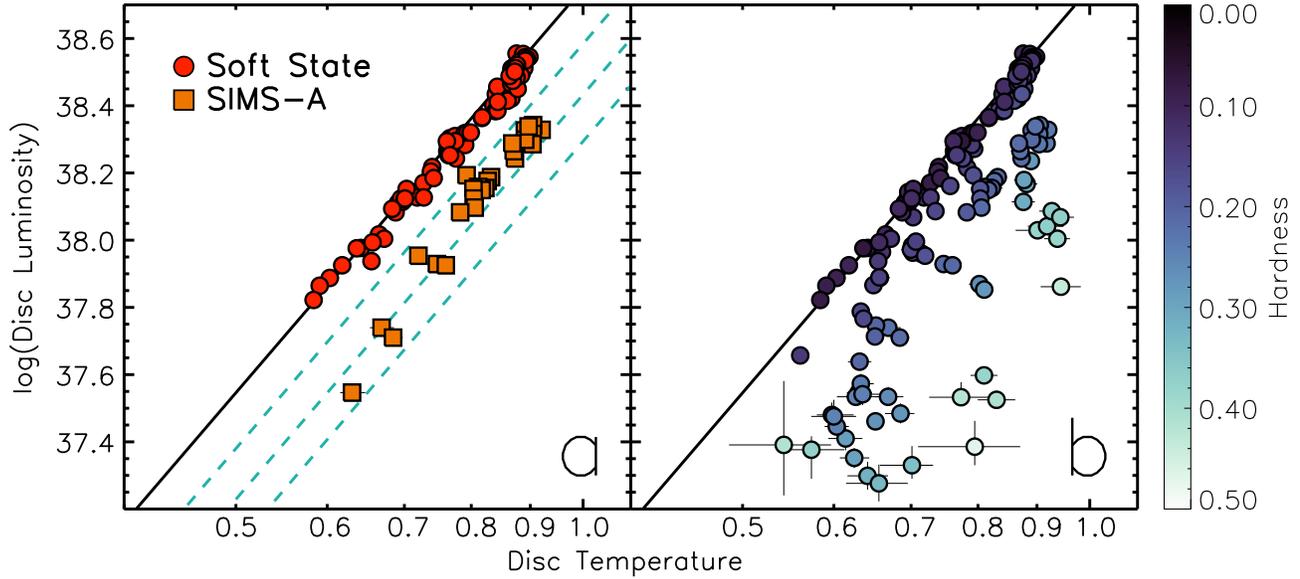, width=\textwidth}
\caption{The inner disc temperature versus unabsorbed bolometric disc luminosity (0.01--1000\,keV). Left: Soft state (red) and SIMS-A (orange) observations. The solid line indicates the $L_{\rm disc} \propto T_{\rm in}^4$ relation which is clearly well fit by the soft state points, while in the SIMS-A the source appears to follow a similar track but at a higher temperature (or lower disc luminosity). The dashed lines indicate increased spectral hardening for factors 1.1, 1.2 and 1.3 larger than the solid line respectively. Right: All disc detections below a hardness of 0.5 are plotted, with the colour scale representing the respective X-ray colour (6--10\,keV / 3--6\,keV) of the observation. The solid line again indicates the $L_{\rm disc} \propto T_{\rm in}^4$ relation revealing there is a clear drop-off with spectral hardness (see also Dunn et al. 2011). Both figures include errors bars but these are typically smaller than the plotted symbol.}
\label{LT4}
\end{figure*}

In Figure \ref{LT4} we plot the observed bolometric disc luminosity versus the inner disc temperature for observations with an X-ray colour ($S_{\rm6-10}/S_{\rm3-6}$) less than 0.5. Throughout all the states this corresponds to $167$ of the $528$ observations used in this study. Figure \ref{LT4}a plots the soft state and SIMS-A observations (red and orange respectively) with the solid line representing the expected $L_{\rm disc} \propto T_{\rm in}^4$ relation and normalised to coincide with the soft state points, displaying a good match (fitted relation: $L_{\rm disc} \propto T_{\rm in}^{\,4.13\pm{0.04}} (1\sigma)$). The SIMS-A observations also appear to follow a similar, albeit marginally steeper, slope (fitted relation: $L_{\rm disc} \propto T_{\rm in}^{\,4.68\pm{0.18}} (1\sigma)$), however they lie on a different plane corresponding to a higher temperature or lower luminosity. The standard SIMS observations fit and even steeper slope of 5.64$\pm${0.19}

We investigated the observations coeval to the SIMS-A and found that generally a combination of both factors led to the change. Since $L_{\rm disc} = 4\pi r_{\rm in}^2\sigma T_{\rm in}^4$ this would in either case require a reduction in the inner radius of the disc, a process that is unlikely given that the soft state is expected to represent the innermost stable circular orbit being reached (see also \citealt{Dunn11}). Furthermore Figure \ref{LT4}b displays each observation by X-ray colour indicating that there is a trend with spectral hardness which continues even into the hard state. A more likely scenario is an increasing $f_{\rm col}$ hence the dotted lines in Figure \ref{LT4}a represent factors 1.1, 1.2 and 1.3 times larger, making the SIMS-A consistent with a moderate modification (see also \citealt{Dunn11}). Curiously however we note that the power-law emission is consistently larger and the reflection fraction generally weaker than the soft state (Figure \ref{PL_R}), even though the SIMS-A observations are distributed fairly evenly among the range of luminosity the soft state exhibits (Figure \ref{HID}). We examine the SIMS-A further in the discussion (\S\ref{SIMS-A}).

\subsubsection{The inner radius from the disc normalisation}\label{norm_d}

\begin{figure}
\centering
\epsfig{file=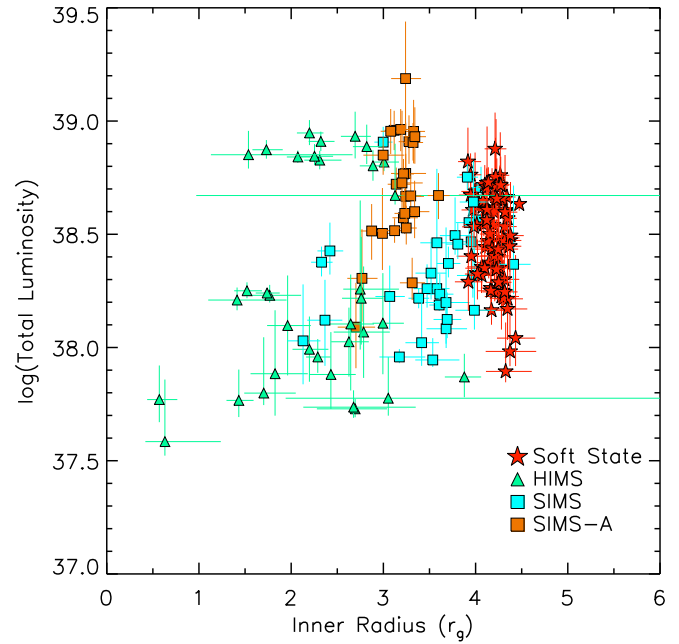, width=0.5\textwidth}
\caption{The disc inner radius, calculated from the \textsc{diskbb} normalisation, versus unabsorbed bolometric disc luminosity (0.01--1000\,keV). The error bars are only statistical, and do not include uncertainties in the mass, distance or inclination of the source.}
\label{rin_nd}
\end{figure}

The disc inner radius can also be calculated from the normalisation of the \textsc{diskbb} model (N$_{\rm D}$):

\begin{equation}
r_{\rm in}/r_{\rm g}=0.677\frac{d_{\rm 10\,kpc}}{M/M_{\sun}}\sqrt{\frac{N_{\rm D}}{\cos{i}}}
\end{equation}
 
In Figure\,\ref{rin_nd} we plot the inferred disc inner radius versus the total source luminosity. In the soft state the inner radius is very stable ($4.19\pm{0.13}\,r_{\rm g}$), however, much like the L--T$^4$ relation, the inner radius appears to be smaller in the intermediate states (see also \citealt{Dunn11}). The SIMS-A points are also consistent with a stable radius, the mean of which is a factor of 1.3 smaller than the soft state, similar to the difference seen in the L--T$^4$ relation (Figure \ref{LT4}). Again, assuming the soft state represents the ISCO, we expect this is due to an increase in the spectral hardening factor rather than the SIMS-A representing a decrease of the inner disc radius. The smaller inner radius may also be an effect of strong Comptonisation not being accounted for (e.g. \citealt{Kubota04}) and we discuss this further in \S\ref{SIMS-A}.

At face value the inner disc radii calculated from the reflection and the disc normalisation disagree, the latter recorded as being considerably smaller. However there are a number of factors that need to be considered. The disc inner radius linearly depends on both the mass of and distance to GX 339$-$4, both of which are highly uncertain. The cosine of the inclination, which is a root term in the equation, is also unknown. The inclination is also required to model the reflection spectrum and the relativistic effects used to measure $r_{\rm in}$, which it can be very degenerate with \citep{Tomsick09,Plant13}. If anything this underlines the current difficulties in precisely measuring the inner radius of the disc and it is hence not surprising that the disc and reflection methods often yield contrasting measurements of black hole spin (see e.g. \citealt{Russell13} for a recent compilation).

%%%%%%%%%%%%%%%%%%%%%%%%%%%%%%%%%%%%%%%%%%%%%%%%%%%%%%%%%%%%%%%%%%%%%%%%%%%%%%%%%%%%%%%%%

\subsection{The ionisation parameter}\label{Xi_1}

\begin{figure}
\centering
\epsfig{file=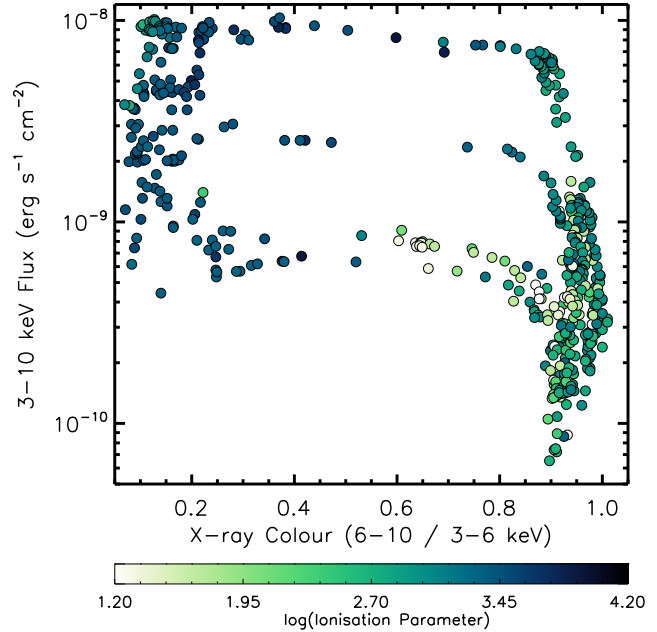, width=0.5\textwidth}
\caption{Evolution of the ionisation parameter throughout the HID. The colour scale represents increased ionisation with darker colours, hence in the soft state the disc is much more ionised. The units of the ionisation parameters are erg\,cm\,s$^{-1}$.}
\label{XI}
\end{figure}

Given the large variation in luminosity as the source progresses through each outburst it is expected that the ionisation of the disc surface layers should vary too, since by definition the ionisation parameter is a function of the flux illuminating the disc. In Figure \ref{XI} we plot how the ionisation parameter evolves throughout the outburst using darker colours to represent increased ionisation. There is a large degree of scatter, however this is not surprising given that this is determined from the soft bandpass and the PCA is limited to $\textgreater$\,3\,keV and in spectral resolution (but see \S\ref{accuracy}). Nevertheless it is easily distinguishable that the soft state generally corresponds to a higher level of ionisation than the hard state.

This is initially quite surprising as the hard state is typically deemed to represent the prime of the Comptonised emission, however as we have shown in Figure \ref{PL_R} the power-law flux remains significant in the soft state. While this can account for the consistently high level of ionisation throughout the outburst it still fails to account for soft state recording higher values. We discuss two explanations for this in \S\ref{Xi_2}: the affect of thermal emission and inner disc truncation. The high level of ionisation maintained throughout the outburst will mean that the reflection albedo is always high and will not vary much \citep{Ross93,Zycki94}.

We also note that at the onset of the hard state during the soft to hard transition the ionisation parameter appears to be at its weakest. This is however the region where our routine can no longer suitably constrain the emission from the accretion disc thus removes the \textsc{diskbb} model. There is however a sufficient amount of thermal emission remaining above 3\,keV at this stage, hence reflection model may be accounting for this excess by reducing the ionisation parameter. We therefore disregard any conclusion we can make from the ionisation parameter during this phase. Once the standard hard track is reached the ionisation returns to a higher level coincident with the rising phase. 

%%%%%%%%%%%%%%%%%%%%%%%%%%%%%%%%%%%%%%%%%%%%%%%%%%%%%%%%%%%%%%%%%%%%%%%%%%%%%%%%%%%%%%%%%

\subsection{Links to compact jets}\label{jets}
In the hard state sources always display steady jet emission at GHz frequencies. Furthermore the X-ray and Radio luminosities appear to be inherently linked, following a non-linear correlation of $L_X\propto L_R^b$, where $0.6\,\textless\,b\,\textless\,0.7$ (\citealt{Corbel03,Gallo03,Corbel13}; but see also \citealt{Coriat11,Gallo12}). Recently \cite{Corbel13} reported a long campaign of quasi-simulatenous X-ray and Radio observations with \emph{RXTE} and the Australia Telescope Compact Array (\emph{ATCA}). This represents the largest sample for a stellar mass black hole, and 32 of these radio observations are coincident (less than 2 days separation) with the data analysed in this study. We therefore investigate how the 8.6\,GHz ATCA observations correlate with the results of our investigation. 

Both the power-law and reflection are well correlated (positively) with the radio luminosity. The power-law correlation coefficient is 0.771, with a false probability of 2.5$\times10^{-5}\,\%$ and best fit slope of $b=0.62\pm{0.01}$, agreeing excellently with \cite{Corbel13}. Since the power-law dominates the total flux in the hard state this agreement is not surprising. The reflection is similarly correlated with a coefficient of 0.664 and a 3.5$\times10^{-3}\,\%$ chance of a false probability. Reflection also reflects a similar relation with the radio luminosity, fitting a slope of $b=0.61\pm{0.02}$. Correlations between the Radio luminosity and other interesting parameters are listed in Table \ref{pRadio}.

\renewcommand{\arraystretch}{1.2}
\begin{table}
\centering
\begin{tabular}{llll}
\hline
Parameter 1		& Parameter 2			& $\rho$		& $p$-value			\\\hline
$S_{\rm radio}$	& $S_{\rm total}$		& 0.765		& 3.38$\times10^{-7}$	\\
$S_{\rm radio}$	& $S_{\rm powerlaw}$	& 0.770		& 2.54$\times10^{-7}$	\\
$S_{\rm radio}$	& $S_{\rm reflection}$	& 0.664		& 3.47$\times10^{-5}$	\\
$S_{\rm radio}$	& Fractional RMS		& $-$0.315	& 0.0793				\\
$S_{\rm radio}$	& $\xi$				& 0.279		& 0.122				\\
$S_{\rm radio}$	& $RF$				& 0.201		& 0.270				\\
$S_{\rm radio}$	& $\Gamma$			& $-$0.013	& 0.946 				\\
\hline	
\end{tabular}
\caption{ A list of SpearmanÕs rank correlations ($\rho$), and the the corresponding false correlation probability ($p$-value), calculated for the 8.6\,GHz radio flux densities listed in Corbel et al. (2013) against fitted parameters in this study. The parameter $S$ refers to the respective flux component.}
\label{pRadio}
\end{table}

%%%%%%%%%%%%%%%%%%%%%%%%%%%%%%%%%%%%%%%%%%%%%%%%%%%%%%%%%%%%%%%%%%%%%%%%%%%%%%%%%%%%%%%%%
%%%%%%%%%%%%%%%%%%%%%%%%%%%%%%%%%%%%%%%%%%%%%%%%%%%%%%%%%%%%%%%%%%%%%%%%%%%%%%%%%%%%%%%%%

\section{Discussion}\label{discussion}
In this work we have utilised three outbursts of GX 339$-$4 to systematically uncover the spectral evolution of black hole binaries. In particular we focus upon the reprocessed reflection spectrum revealing marked changes in the reflection, often not in tandem with the Comptonised emission. As we will discuss now, the most likely explanation for this is distinct evolution of the accretion geometry.

%%%%%%%%%%%%%%%%%%%%%%%%%%%%%%%%%%%%%%%%%%%%%%%%%%%%%%%%%%%%%%%%%%%%%%%%%%%%%%%%%%%%%%%%%

\subsection{The overall picture}\label{picture}
\begin{figure*}
\centering
\epsfig{file=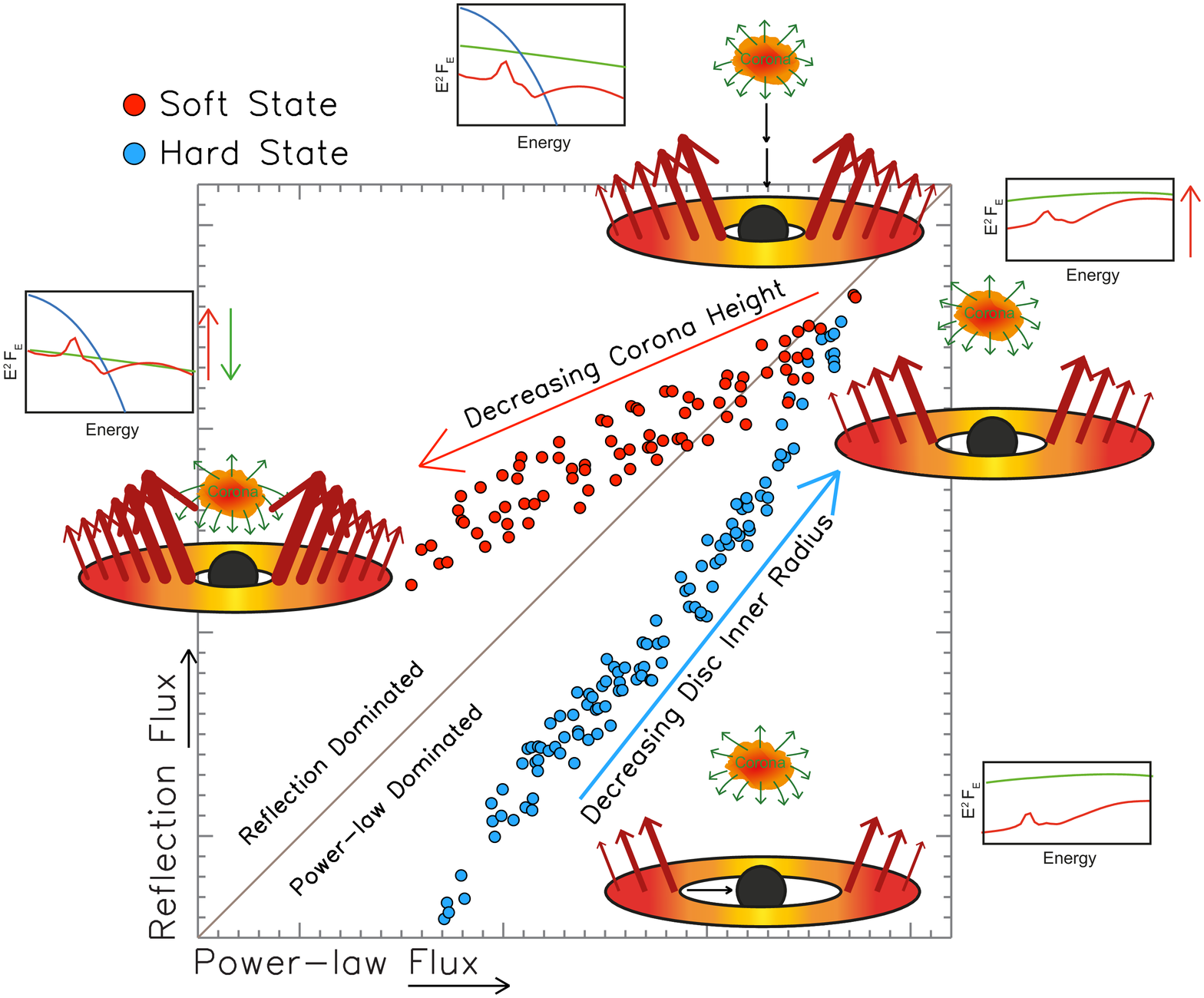,  width=\textwidth}
\caption{An illustration of how geometrical evolution will lead to contrasting changes in the Comptonised and reprocessed emission, and notably an increase in reflection fraction. Hard State (blue): The reflection fraction can be increased by decreasing the inner radius of the accretion disc. For a stable corona this model increases the solid angle subtended by the disc, thus increasing the photons intercepted and reprocessed flux. The Comptonised emission meanwhile remains constant. This interpretation suits the hard state well, where the reflection fraction is low, consistent with the small solid angle of a truncated disc, but gradually increases as the source rises. Soft State: If the scale height of the corona decreases, the portion of Comptonised photons intercepted by the accretion disc will increase, leading to increased reprocessing and thus reflection flux observed at infinity. In addition, the lower height will lead to stronger relativistic effects, ultimately focusing more Comptonised photons onto the disc, and in particular the inner regions. As well as heightened reflection, the Comptonised flux will also decrease as a consequence of the increased anisotropy of the emission. Similar results can be formed with alternative modifications of the corona that ultimately lead to irradiation from a lower mean height. We discuss these models in more details in \S\ref{picture}.}
\label{BH}
\end{figure*}

In Figure \ref{PL_R} we display how the Comptonised and reflection emission co-evolve as the outburst progresses. In the hard state (1) the power-law is typically responsible for 5 times more flux than the reflection ($RF\sim0.2$), however as the source rises the reflection fraction increases yet always remains below $1$. On the other hand, the soft state lies in the main above unity, thus signifying larger levels of reflection than power-law. Furthermore, as the source decays the reflection fraction appears to continue increasing. What makes this interesting is that for a stable geometry the two should vary in tandem as the Comptonised emission feeds the reprocessing observed as the reflection spectrum. Therefore evolution is a signal of changes in the accretion geometry.

Figure \ref{BH} illustrates two such transformations that can lead to the observed cycle portrayed in Figure \ref{PL_R}. The truncated disc model has become a popular means to describe the observed characteristics of the hard state, such as the hard spectrum and low reflection fraction (see \citealt{Done07} for a review and references therein). As the inner radius of the accretion decreases the solid angle it subtends beneath the illuminating corona increases. More photons are thus intercepted leading to increased reprocessing and reflected flux. In turn the amount of soft photons entering the corona will increase as the disc moves further towards the black hole leading to cooling of the hot electrons, which is observed through a softer photon index and high energy cut-off (Figures \ref{gamma} and \ref{cutoff}).

\cite{Beloborodov99} proposes an alternative explanation for these trends through a `dynamic corona', whereby bulk motion of the emitting plasma reduces the irradiation of the disc. Even with a disc at the ISCO this regime can yield the small reflection fractions observed in the hard state, whilst the beaming will also suppress the soft seed photons, hence retaining a hard spectrum ($\Gamma\sim1.6$). We however observe that both $RF$ and $\Gamma$ increase as the source rises through the hard and into the intermediate states (see \S\ref{PL} and \S\ref{RF}), which would thus require a decreasing bulk motion. This would appear at odds with the favoured internal shocks model outlined in \cite{Fender04}, whereby the jet velocity increases with luminosity/state. Given that there is no deviation from the observed trends in $RF$ and $\Gamma$ would suggest that the illuminating medium remains consistent as well. Furthermore the same process ruling $RF$ and $\Gamma$ is present in the decay of the hard state whereby they both decrease as the source fades ($RF$ fitted slope $1.21\pm{0.06}$), thus in the dynamic corona regime the jet velocity should systematically increase leading to bright radio flares. The reality however is that such flares have only been observed in the rise (\citealt{Fender04}, see also \citealt{Corbel13b}), thus we therefore propose the truncated disc model as our favoured interpretation. High resolution observations with \emph{XMM-Newton} and \emph{Suzaku} have also confirmed that the inner disc radius is truncated and decreases with increasing luminosity in the hard state (\citealt{Plant13,Kolehmainen13}; see also \citealt{Petrucci14}). The lower values for the ionisation parameter recorded in the hard state also agree with a truncated inner disc (see \S\ref{Xi_2}). Furthermore, the reflection component correlates better with the reflection fraction than the Comptonised one (Table \ref{pFlux}), hence this would suggest the changes are driven by the geometry of the disc changing, rather than the corona. The reflection fraction is also almost identical both in magnitude and evolution during the rise and decay (stages 1 and 4) of the hard state (compare Figures \ref{PL_R}b--c) which argues strongly for an identical dynamic in and out of quiescence.

In contrast, the reflection dominates the Comptonised emission in almost all of the soft state observations ($RF\,\textgreater1$). Furthermore, as the decay progresses the reflection fraction appears to increase. The scatter in flux due to the reduced hard band signal in the soft state makes fitting the trend difficult, however if the 2002 outburst is excluded a slope of $\alpha=0.49\pm{0.17}$ is resolved (where $\log_{10} S_{\rm PL}=\alpha\log_{10} S_{\rm Ref}+c$). In addition, fitting the SIMS and SIMS-A observations, which track the decay quite well (Figure \ref{HID}), reveals a similar relationship ($\alpha=0.73\pm{0.07}$). The decay of the disc luminosity fits the expected $L_{\rm disc} \propto T_{\rm in}^4$ relation well, hence the inner disc radius is without any reasonable doubt stable in the soft state (see \S\ref{L-T} and also \citealt{Gierlinski04,Steiner10,Dunn11}), thus we can rule out the same mechanism as the hard state. Instead, this acts as a strong indication that the underlying changes modifying the reflection fraction are occurring in the corona. The Comptonsied component correlates better with the reflection fraction than the reflection in the soft and soft-intermediate states, and hence acts as further evidence for this dynamic.

In Figure \ref{BH} we illustrate the relationship between the coronal height and the amount of reflection and Comptonised flux observed. A reduction in height increases the portion of Comptonised photons intercepted by the accretion disc, leading to increased reprocessing and thus reflection flux observed at infinity. In addition the lower height will lead to stronger relativistic effects upon each Comptonised photon, ultimately focusing more onto the disc, and in particular the inner regions \citep{Miniutti04,Wilkins12}. This will firstly heighten the reflection flux further due to the increased irradiation. Secondly the Comptonised flux observed will decrease as a result of the increased anisotropy (and tolerance towards the black hole) of the Comptonised emission with lowering height. It is clear to see then how the coronal height can significantly vary the reflection fraction by coupling enhanced reflection with a diminished power-law and we therefore favour this solution as the process behind the increasing reflection fraction in the soft state. While we have showcased the corona as a compact `lamp-post' for illustration purposes, the effect of the coronal height may manifest itself in a different form, such as by varying in extension or collapsing, however the interpretation is still very much alike: the coronal geometry is expected to be evolving as the source decays in the soft state. 

By using the evolving $RF$ to constrain the accretion geometry we have assumed that the albedo remains constant. This however may not be true since the surface layers of the disc are subjected to a large range of illumination during the different phases of the outburst. The albedo itself is strongly dependant upon the ionisation of the surface layers, hence it must be considered how this may influence our results. Above 10\,keV the scattering cross-section exceeds the absorption cross-section, hence there the albedo of the reflection is close to unity and constant. However, below 10\,keV the albedo is strongly influenced by the ionisation stage: as the layers become more ionised the absorptive opacity decreases, leading to more effective reflection \citep{Ross93, Matt93, Zycki94}. Ultimately a fully ionised layer should act like mirror, corresponding to an albedo of unity.

In the case of BHXRBs the surface layers are expected to be kept highly ionised by the hot thermal emission emerging from the disc itself \citep{Ross93}, which is evident by the high ionisation parameter values we record throughout this study (see \S\ref{Xi_1}). Furthermore, the ionisation values we record are large enough that the albedo should remain consistently high, and not far from unity \citep{Zycki94}. We do see some evolution in the ionisation parameter, which particularly contrasting between the hard and soft states (Figure \ref{XI}), but this is not enough to change the albedo by more than a factor of 2 \citep{Zycki94}. As we have described previously, the reflection fraction varies over a much broader range than this in outburst.

%%%%%%%%%%%%%%%%%%%%%%%%%%%%%%%%%%%%%%%%%%%%%%%%%%%%%%%%%%%%%%%%%%%%%%%%%%%%%%%%%%%%%%%%%

\subsection{On the nature of the SIMS-A observations}\label{SIMS-A}

When classifying spectral states it became apparent that there was two distinct clusters of points with low variability (RMS$\textless5\%$) isolated by their X-ray colour (6--10/3--6\,keV flux; Figure \ref{RMS_hard}). We associate the softer points (X-ray colour $\textless0.175$) with the canonical soft state, whilst the harder points are described by spectra very similar to the SIMS. Additionally, of the 7 observations containing a type-A QPO listed in \cite{Motta11}, 6 are in this group. The seventh observation covered what appears to be part of a transition, hence only a portion of the time-series was used by \cite{Motta11} meaning a RMS of $\sim2\%$ was reported. In our study we utilised the entire observation which raised the RMS to $\sim6\%$. As a result of this association we labelled these points as SIMS-A. The majority of the SIMS-A observations displaying a type-A QPO are close to the top branch of the HID (Figure \ref{HID}), and given that the feature is weak and broad \citep{Motta11} it is likely that many SIMS-A observations where a QPO was not detected simply lack the signal to do so.

As stated before the SIMS-A has a very similar spectrum to the SIMS (Table \ref{avpar}). In the HID (Figure \ref{HID}) the SIMS-A appear to be distributed at a slightly higher luminosity than the SIMS, however this distinction becomes more clear in Figure \ref{PL_R}d. The SIMS-A is therefore probably associated with an increase in the Comptonised and reflection components, and is the likely explanation for the hardened X-ray colour. It is also interesting that the SIMS-A remains on the `soft' slope of $\textless1$ (Figure \ref{PL_R}d) describing the general decay of the source, even when there is an apparent increase in power-law and reflection flux each time the source transitions to it.  

\begin{figure}
\centering
\epsfig{file=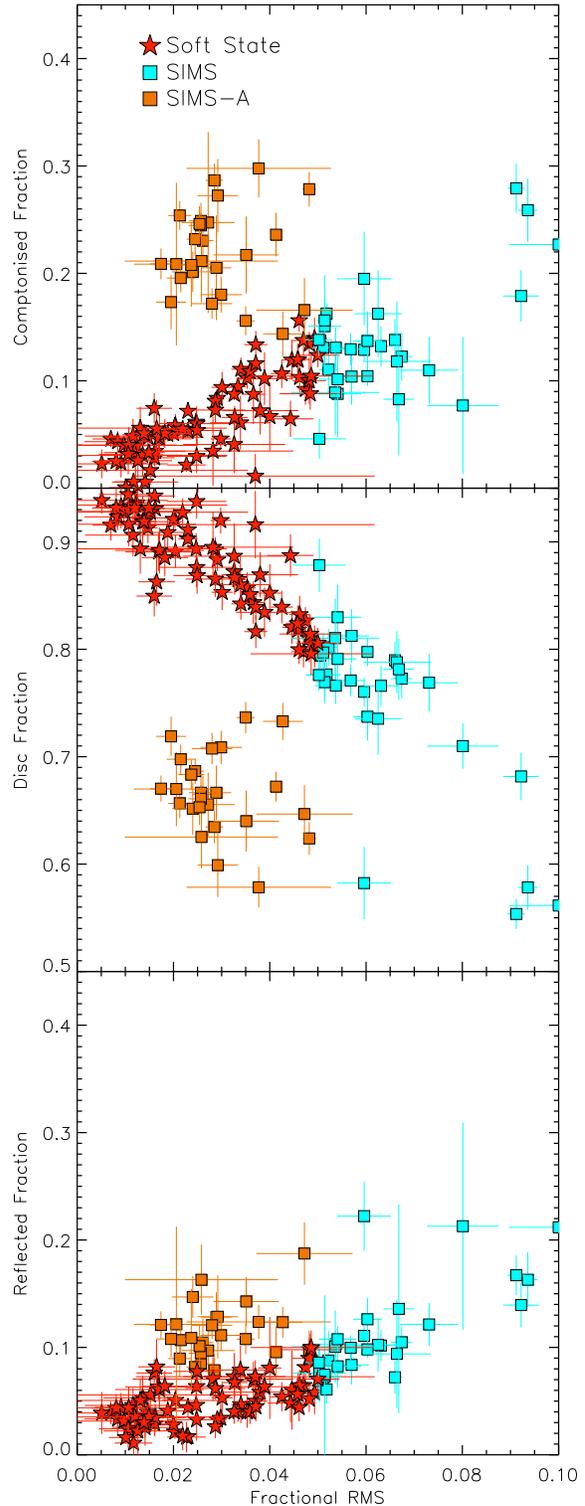, width=0.42\textwidth}
\caption{The fraction of total source flux that comes from the Comptonised (top figure), disc (middle) and reflection (bottom) components respectively. We plot the SIMS, SIMS-A and soft state observations to indicate how the source spectrum is different between these three principle states. Whilst the fractional RMS has proven to be a better indicator of state than spectral hardness, clearly the states are not completely distinct, particularly between the SIMS and soft state (see \S\ref{states}). Note that the plotted reflected fraction is the ratio of the reflection and total source flux, and should not be confused with the $RF$. We calculated each flux in the same band as the RMS (2--15\,keV) to make the two axis directly comparable, however the same trend is seen using 0.1--1000\,keV flux as well.}
\label{Frac_RMS}
\end{figure}

The spectrum of the SIMS-A is dominated by the power-law and reflection, whereas the disc flux is relatively weak (Table \ref{avpar}). In comparison, the disc generally accounts for the majority of flux in the SIMS, as is of course standard for the soft state as well. In Figure \ref{Frac_RMS} we plot the fraction of the total source flux from each component which also emphasises how the disc and Comptonised flux are decreased and increased respectively in the SIMS-A.

In total there are 13 instances of transitions to the SIMS-A recorded in our study (i.e. excluding observations already in the SIMS-A). Six of these transitioned from the SIMS, of which 5 lead to an increase in reflection flux and 4 in power-law flux. In 4 of these the disc flux also decreased. Another six observations transitioned from the soft state, all of which are characterised by increases in power-law and reflection flux in addition to a reduced disc flux. Furthermore, in each case the photon index softened and all but one of these transitions lead to a net increase in flux. In the remaining observation the source transitioned directly from the HIMS. It is also interesting that the Comptonised and disc fractions plotted in Figure \ref{Frac_RMS} isolate the SIMS-A observations quite well, whereas the SIMS appears more to be an extension of the soft state. This is typically true for SIMS observations with RMS in the 5--7\,\% range, whilst most of the type-B QPOs reported in \cite{Motta11} correspond to observations with a RMS above this level. We note that the choice of the 5\,\% RMS line separating the SIMS and soft state is rather arbitrary and the true value could be slightly higher. This region is also prone to fast transitions shorter than the typical exposure time of the PCA \citep{Munoz11}.

Interestingly, the SIMS-A tend to roughly follow the L$\propto$T$^4$ relation (fitted relation: $L_{\rm disc} \propto T_{\rm in}^{\,4.68\pm{0.18}} (1\sigma)$) commonly attributed to the soft state, but at a lower luminosity (or higher disc temperature), consistent with an spectral hardening factor 1.1--1.3 times larger (Figure \ref{L-T}). An alternative explanation is a reduced disc inner radius, however this is unlikely given that the soft state is well regarded to harbour a disc extending all the way to the ISCO. The inner radius derived from the disc normalisation is also fairly stable and less than that of the soft state. The position on the L--T plot and smaller inner radii are both consistent with the \emph{anomalous} regime of GRO J1655$-$40 \citep{Kubota01} and XTE J1550$-$564 \citep{Kubota04}.

In relation to the soft state it is quite easy to explain the harder X-ray colour in the SIMS-A, simply since there is a decrease in the soft (disc) contribution whilst the hard (power-law/reflection) increases. However this fails to explain the consistently low level of variability usually associated with emission from the disc (or lack of Comptonised emission). The SIMS-A do appear to show a similar relationship between the Comptonised fraction and RMS as seen for the SIMS and soft state observations, but offset to a higher fraction. It may be that an additional non-variable component is also present but compensated for by the \textsc{cutoffpl} model. Curiously, in the \emph{anomalous} regime reported by \cite{Kubota01} and \cite{Kubota04} an extra component is required, which is harder than the disc but softer than the power-law, and when accounted for returns the regime to the soft state L--T$^4$ track. This may account for the disparity between the inner radii derived from the reflection and the disc. The same process giving rise to the type-A QPO may be behind the increased hardening of the thermal emission or Comptonised flux as well. We note that the SIMS-A are nevertheless well fitted (Figure \ref{redchi}).

%%%%%%%%%%%%%%%%%%%%%%%%%%%%%%%%%%%%%%%%%%%%%%%%%%%%%%%%%%%%%%%%%%%%%%%%%%%%%%%%%%%%%%%%%

\subsection{The accuracy of this study and comparisons to other works}\label{accuracy}
\begin{figure}
\centering
\epsfig{file=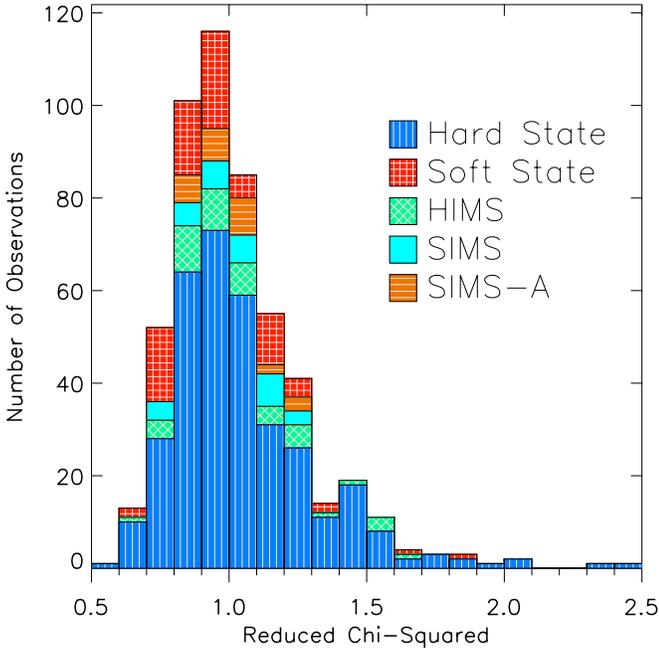, width=0.5\textwidth}
\caption{The reduced-$\chi^2$ distribution for each best-fit model in our analysis, with 95\% of observations reporting $\leq1.5$. }
\label{redchi}
\end{figure}

In using \emph{RXTE} our approach is quantitative more than qualitative, hence it is important to examine the accuracy of our study. To do this we can examine how well investigations with high-resolution missions like \emph{XMM-Newton} and \emph{Suzaku} agree with our results. Such studies of the hard state \citep{Reis10,Done10,Plant13,Kolehmainen13} reveal moderate ionisation ($\textless\log{3}$) and a low reflection fraction akin to our results. Studies of the HIMS are sparse due to the short duration of this state, but observations analysed by \cite{Hiemstra11} and \cite{Reis11,Reis12} agree with the evolution outlined in \S\ref{spectra}, in particular an increased level of ionisation ($\textgreater\log{3}$) and reflection fraction (the equivalent width is now 250--450\,keV vs. 50--150\,keV in the hard state). \cite{Kolehmainen11} investigated the SIMS and soft state of GX 339$-$4, confirming consistently high ionisation and increased reflection fraction habitual to the soft state in this study. Furthermore, they discuss difficulties in fitting the Fe line profile in the soft state, which is hampered since the disc is providing most of the flux and additional curvature in the continuum at 6\,keV. This is most likely reason for the large scatter in $r_{\rm in}$ during the soft state (Table \ref{avpar}).

We can also compare our results to previous works with \emph{RXTE}. Recently \cite{Reis13} performed a similar analysis with one outburst of the black hole XTE J1650$-$500 finding similar results. For example, they also plot reflection versus power-law flux displaying similar evolution throughout the states (Figure \ref{PL_R}). \cite{Dunn08} also studied reflection in GX 339$-$4, instead fitting the Fe line with a Gaussian. Remarkably, even with this simpler approach, very similar evolution is revealed. In their Figure 8 \cite{Dunn08} plot Fe line flux against 7--20\,keV flux (essentially reflection versus power-law flux) finding the same looped evolution and an increased fraction of reflection to power-law in the softer states. Furthermore, the fitted slopes to their `hard' and `soft' observations are almost identical to those in this study (Figure \ref{PL_R}b--d), whilst the evolution of the equivalent width in their Figure 7 is very similar to the trend of its analogous parameter $RF$ analysed in this investigation (Figure \ref{PL_R}).

It is also possible that the three components (disc, power-law and reflection) exclusively assumed are not the full description of the X-ray spectra from BHXRBs. For example, in the AGN community much controversy exists due to the potential effect of absorption, and in particular the level, even in some cases presence, of reflection \citep{MillerL08}. No strong indicator of absorption, such as dips, eclipses or narrow lines has been observed from this source \citep{Ponti12}, although this does not necessarily rule out that absorption is present or influences our results. It should be noted though that the reflection spectrum will also be influenced by photons emitted in the disc \citep{Ross93,Ross07}, and this is not currently accounted for by \textsc{xillver-a} and is most likely to have an effect in the soft state where the disc is very strong. 

The convolution model \textsc{rfxconv}, which combines the ionised reflection of \textsc{reflionx} \citep{Ross05} and the angle-dependant Compton reflection of \textsc{pexrav} \citep{Magdziarz95}, can accept any input continuum, such as that from a disc. \cite{Kolehmainen11} applied \textsc{rfxconv} to soft and soft-intermediate state \emph{XMM-Newton} observations of GX 339$-$4, which through the soft bandpass could directly infer the disc contribution to the reflection spectrum. The results of \cite{Kolehmainen11} are very similar to ours (see e.g. their Table 3), hence we believe our analysis using \textsc{xillver-a} should be robust to the effects of the disc contribution. However, we do note that self-consistent ionised reflection from illumination by the Comptonised \emph{and} disc emission is needed to confirm our results. Finally the chi-square distribution is excellent, with 95\% of observations having a reduced--$\chi^2$ of $\leq1.5$ (Figure \ref{redchi}).

%%%%%%%%%%%%%%%%%%%%%%%%%%%%%%%%%%%%%%%%%%%%%%%%%%%%%%%%%%%%%%%%%%%%%%%%%%%%%%%%%%%%%%%%%

\subsection{Why is the reflection spectrum more ionised in the soft state?}\label{Xi_2}

In \S\ref{Xi_1} we analyse the ionisation parameter, defined as the ratio of the illuminating flux and the gas density, where it emerges that the parameter is significant larger in the soft state. This would appear at odds with the state of the illuminating flux, which typically peaks in the brighter stages of the hard state and subsequent transition into the HIMS (Figure \ref{PL_R}). To explain this one important aspect to consider is the increased disc emission in the soft state, which will undoubtedly have an ionising affect on the surface layers in addition to the illumination from the corona above. To this end \cite{Ross07} investigated the impact of the disc, finding that increased thermal radiation and peak temperature will ultimately result in a more ionised spectrum. In particular the Fe profile is strongly affected as a result of higher ionisation stages and greater Compton-broadening. Our chosen reflection model (\textsc{xillver-a}) does not account for change in the thermal emission throughout the outburst and thus is likely to react to changes in the disc by varying the ionisation parameter, in particular with an increase in softer states.

Another explanation for the apparent change in illumination could result from varying the area of the disc and thus the solid angle it subtends below the corona. If the inner accretion disc is truncated then the amount of Comptonised photons intercepted by the disc will decrease rapidly. Furthermore, assuming a lamppost geometry, the illumination pattern should roughly go as R$^{-3}$, thus the reprocessed spectrum will be dominated by emission from the most central region, which in turn likely represents the most ionised zone of the disc because of the peaked illumination there. Even with a small level of truncation the inferred ionisation level will probably diminish significantly, and this contrast may also be heightened if light-bending is at play (see Figure \ref{BH}). This argument works well since, as discussed in \S\ref{accuracy}, high-resolution spectroscopy resolves the inner disc to be truncated and of a lower ionisation stage in the hard state (e.g. \citealt{Plant13}).

%%%%%%%%%%%%%%%%%%%%%%%%%%%%%%%%%%%%%%%%%%%%%%%%%%%%%%%%%%%%%%%%%%%%%%%%%%%%%%%%%%%%%%%%%
%%%%%%%%%%%%%%%%%%%%%%%%%%%%%%%%%%%%%%%%%%%%%%%%%%%%%%%%%%%%%%%%%%%%%%%%%%%%%%%%%%%%%%%%%

\section{Summary}
In this study we have performed a comprehensive and systematic investigation of X-ray reflection from GX 339$-$4. In total we analysed 528 observations made by \emph{RXTE}, covering the three full outbursts between 2002 and 2007. This represents the largest study of X-ray reflection applying a self-consistent treatment to date, and in particular such excellent monitoring has allowed a thorough investigation of state transitions, in addition to the canonical hard and soft states (Figure \ref{HID}). Each observation was well described by a combination of thermal disc emission (\textsc{diskbb}), a cut-off power-law (\textsc{cutoffpl}) and relativistically blurred reflection (\textsc{relconv$\ast$xillver-a}). All hard state and some HIMS observations did not require a disc due to the 3\,keV lower limit to the PCA bandpass, while the bright stages of the hard state display a decreasing high-energy cut-off as the source luminosity rises (Figure \ref{cutoff}). X-ray reflection is required for the entire duration of the study.

We pay particular detail to how the power-law and reflection co-evolve throughout the outburst which display a very strong positive correlation throughout each state (Table \ref{pFlux}). Since the reflection arises as a consequence of the power-law irradiating the disc, contrasting evolution acts a strong indicator of geometrical changes. The hard state is distinctly reflection weak with a typical reflection fraction (the ratio of reflection and power-law flux) of $\sim$\,0.2. The reflection fraction does however increase as the source luminosity rises fitted by a slope of $\sim$\,1.2 and $\sim$\,1.7 in the fainter and brighter stages of the hard state rise respectively (Figures \ref{PL_R}a and \ref{PL_R}b). The latter indicates a change in the dynamic behind the increasing reflection fraction and occurs during the phase where the high-energy cut-off in the Comptonised emission decreases (Figure \ref{cutoff}). In stark contrast the soft state represents a period of strong reflection whereby the reflection fraction exceeds unity for almost the entire state. Furthermore, while there is some scatter, the fitted slope to the 2004 and 2007 outbursts is $0.49\pm0.17$, signifying a furthering of the reflection dominance over the power-law emission. As the source decays into the hard state the reflection fraction returns to the same magnitude and slope as in the rise, hence whether the source heading towards or out of quiescence the accretion geometry is likely to be the same.

We discuss what is driving the contrast between the reflection and power-law in \S\ref{picture}. The favoured interpretation is a truncated inner disc radius and decreasing illuminating source height for the hard and soft states respectively (see Figure \ref{BH} for an illustration). The hard state inner accretion disc of GX 339$-$4 was recently found to be truncated and decrease in radius with increasing luminosity \citep{Plant13}, which explains qualitatively how the low reflection fraction can be achieved whilst gradually promoting its increase. Softening of the power-law photon index (\S\ref{PL}), a decreasing high-energy cut-off, and a low ionisation parameter (\S\ref{Xi_1} and \S\ref{Xi_2}) all add strength to the truncated disc interpretation. The decay in disc flux and temperature should directly follow that of a black-body for a constant disc area ($S_{\rm disc}\propto T^4$). The soft state tracks a very similar relation of $T^{4.13\pm{0.04}} (1\sigma$; Figure \ref{LT4}) implying that the inner disc radius is constant and at the ISCO. This immediately explains how the reflection fraction is able to reach unity (essentially a solid angle of $2\pi$) and confines geometrical changes to the corona. We interpret the increase in reflection fraction as the source decays through a decreasing corona height strengthening the disc illumination and the effects of light-bending.

%%%%%%%%%%%%%%%%%%%%%%%%%%%%%%%%%%%%%%%%%%%%%%%%%%%%%%%%%%%%%%%%%%%%%%%%%%%%%%%%%%%%%%%%%

\section*{Acknowledgements}
The authors would like to thank Mari Kolehmainen for helpful discussions. DSP acknowledges financial support from the STFC. GP acknowledges support via an EU Marie Curie Intra-European Fellowship under contract no. FP-PEOPLE-2012-IEF-331095. TMD acknowledges funding via an EU Marie Curie Intra-European Fellowship under contract no. 2011-301355. Part of this work was supported by the COST Action MP0905 ``Black Holes in a Violent Universe" and DSP also thanks the Max Planck Institute fur Extraterrestriche Physik for their hospitality during a short visit. Many figures in this work have made use of the cubehelix color-scheme \citep{Green11}.
\bibliographystyle{mn2e_fix}
\bibliography{refs}

\end{document}